\documentclass[a4paper,fleqn,usenatbib,useAMS]{mnras}

\usepackage{graphicx}	
\usepackage{amsmath}	
\usepackage{amssymb}	
\usepackage{multicol}        

\usepackage{bm}		
\usepackage{pdflscape}	
\usepackage{xcolor}

\usepackage[T1]{fontenc}
\usepackage{ae,aecompl}

\usepackage{newtxtext,newtxmath}

\title[A Census of LyC Photons]{A Census of the LyC Photons that Form the UV Background During Reionization}

\author[H. Katz]{Harley Katz$^{1,2}$\thanks{Contact e-mail: \href{mailto:harley.katz@physics.ox.ac.uk}{harley.katz@physics.ox.ac.uk}}, Taysun Kimm$^{2,3}$, Martin Haehnelt$^2$, Debora Sijacki$^2$, Joakim Rosdahl$^4$, \newauthor and Jeremy Blaizot$^4$
\\
$^{1}$Astrophysics, University of Oxford, Denys Wilkinson Building, Keble Road, Oxford OX1 3RH, UK\\
$^{2}$Institute of Astronomy and Kavli Institute for Cosmology, Cambridge, Madingly Road, Cambridge, CB3 0HA, UK\\
$^{3}$Department of Astronomy, Yonsei University, 50 Yonsei-ro, Seodaemun-gu, Seoul 03722, Republic of Korea\\
$^{4}$Univ Lyon, Univ Lyon1, ENS de Lyon, CNRS, Centre de Recherche Astrophysique de Lyon UMR5574,
F-69230, Saint-Genis-Laval, France}

\date{\today}

\pubyear{2017}

\begin{document}
\label{firstpage}
\pagerange{\pageref{firstpage}--\pageref{lastpage}}
\maketitle

\begin{abstract}
We present a new, on-the-fly photon flux and absorption tracer algorithm designed to directly measure the contribution of different source populations to the metagalactic UV background and to the ionisation fraction of gas in the Universe. We use a suite of multifrequency radiation hydrodynamics simulations that are carefully calibrated to reproduce a realistic reionization history and galaxy properties at $z \ge 6$, to disentangle the contribution of photons emitted by different mass haloes and by stars with different metallicities and ages to the UV background during reionization. While at very early cosmic times low mass, metal poor haloes provide most of the LyC photons, their contribution decreases steadily with time. At $z = 6$ it is the photons emitted by massive systems (${\rm M_{halo}}/{\rm M_\odot} > 10^{10} \, {\rm h ^{-1}}$) and by the metal enriched stars ($10^{-3} < Z/Z_{\rm \odot} < 10^{-1.5}$) that provide the largest contribution to the ionising UV background. We demonstrate that there are large variations in the escape fraction depending on the source, with the escape fraction being highest ($\sim 45-60\%$) for photons emitted by the oldest stars that penetrate into the IGM via low opacity channels carved by the ionising photons and supernova from younger stars. Before HII regions begin to overlap, the photoionisation rate strongly fluctuates between different, isolated HII bubbles, depending on the embedded ionising source, which we suggest may result in spatial variations in the properties of dwarf galaxies.
\end{abstract}




\section{Introduction}
Various observational techniques, such as measuring the Gunn-Peterson optical depth from QSO spectra or the prevalence of Ly$\alpha$ emission in high redshift galaxies, have placed very tight constraints on the volume filling fraction of neutral hydrogen in the intergalactic medium (IGM) towards the end of reionization at $z\sim6$ \citep{Fan2006,McGreer2015,Schroeder2013,Totani2006,McQuinn2008,Ouchi2010,McQuinn2007,Ota2008,Caruana2014,Ono2012,Sobacchi2015,Mesinger2015,Mortlock2011,Bolton2011,Tilvi2014,Schenker2014,Pentericci2014,Becker2013,Robertson2013,Becker2001,Chornock2013,Mitra2015,Mitra2016,Greig2017}.  While the timing of the end of reionization is rather well constrained by observations and modelling \citep[e.g.,][]{Fan2006,Choudhury2015} as is the photoionisation rate of neutral hydrogen in the post-reionization Universe \citep{Bolton2007,Wyithe2011,Calverley2011,Becker2013}, much remains uncertain about the onset and extent of the process.  This uncertainty is primarily driven by the current lack of understanding of which sources reionized the Universe and the difficulty of observing these systems deep into the epoch of reionization.  Various classes of objects have been proposed as the sources of reionization including dwarf galaxies, mini-haloes, massive galaxies, active galactic nuclei, accretion shocks, globular clusters, stellar mass black holes, and dark matter annihilation and decay \citep{Madau2015,Couchman1986,Madau1999,Shapiro1987,Haiman1998,Dopita2011,Ricotti2002,Katz2013,Katz2014,Madau2004,Ricotti2004,Mirabel2011,Mapelli2006}.  

Differentiating between the possible sources of reionization requires an accurate determination of both the number density of the sources as well as the escape fraction of Lyman continuum (LyC) photons that penetrate into the IGM.  If one extrapolates the observed galaxy UV luminosity functions to some limiting magnitude and converts this into a star formation rate (SFR) as a function of redshift, the ionising photon contribution from galaxies can be determined as
\begin{equation}
\dot{n}_{\rm ion}=\xi_{\rm ion}f_{\rm esc}\rho_{\rm SFR},
\label{eqn1}
\end{equation}
where $\dot{n}_{\rm ion}$ is the number density of ionising photons per unit time that are injected into the IGM, $\xi_{\rm ion}$ is the number of ionising photons emitted per second per unit SFR, which can be calculated assuming a stellar initial mass function (IMF), $\rho_{\rm SFR}$ is the star formation rate density, and $f_{\rm esc}$ is the escape fraction of ionising photons that penetrate out from the galaxy and into the IGM \citep[e.g.,][]{Robertson2015}.

The main uncertainty in this equation is the value of $f_{\rm esc}$.  Observational determinations of $f_{\rm esc}$ are extremely difficult.  For galaxies at $z\lesssim1$, estimates of $f_{\rm esc}$ tend to be lower than what is required for reionization \citep{Siana2007,Siana2010,Bridge2010,Leitet2011,Leitet2013,Rutkowski2016,Leitherer2016} although, certain lower metallicity galaxies with high SFRs may exhibit higher values \citep{Borthakur2014,Izotov2016a,Izotov2016b}.  This measurement becomes even more challenging at higher redshifts \citep[e.g.,][]{Mostardi2015,Reddy2016}.  Due to these observational challenges, numerical simulations have become an invaluable tool for measuring $f_{\rm esc}$ \citep{Gnedin2008,Wise2009,Razoumov2010,Yajima2011,Paar2013,Kimm2014,Wise2014,Paar2015,Ma2015,Xu2016,Trebitsch2017,Kimm2017,Rosdahl2018}; however, disagreements between different simulations persist.

Unfortunately, measuring the escape of ionising photons from a galaxy and applying Equation~\ref{eqn1} does not provide a full picture of which sources ionise the Universe by mass and by volume, nor does it measure the contribution to the metagalactic UV background\footnote{In the post-reionization Universe, when the IGM is optically thin, we expect that the contribution to the UV background by each different class of source can be predicted by knowing the star formation rate and $f_{\rm esc}$ since absorptions in the IGM are likely negligible.}.  Environment plays a very important role since a galaxy residing in a much lower density region can ionise more volume than the same galaxy forming in a deep potential well due to the drastic differences in the number of recombinations.  Likewise, clustering, merging, virialization redshift, and neighbour properties all impact what percentage of emitted photons reach the lower density IGM.  All of these types of effects can only be captured in a full cosmological simulation that follows the history of the emission and absorption of photons from various types of sources. 

In this paper, we present a new algorithm where we can track the sources of individual photons in order to overcome these major difficulties in identifying what role each individual class of sources plays in reionizing the Universe, both in terms of the contribution to the metagalactic UV background and to volume-weighted and mass-weighted ionisation fractions.  This crucially includes sources that can be hosted by the same object; however, the algorithm is equally useful for tracking global galaxy properties.  The method we present  bypasses the need to measure escape fractions for individual sources and directly measures the contribution of each source to the global photoionisation rate at every given epoch in the simulation as well as the fraction of the IGM ionised by each of the different classes of sources.  However, we show that the algorithm can also be used to calculate the globally averaged escape fraction of each source class present within the simulation at a time resolution of the smallest hydrodynamic time step.  We present multiple demonstrative cases where we track LyC photons based on halo mass, stellar metallicity, and stellar age to determine which photons actually leak into the IGM and drive reionization.

This paper is organised as follows.  In Section~\ref{ptalg}, we introduce our photon flux and absorption tracer algorithm that has been implemented into a moment based, multifrequency, on-the-fly radiative transfer code.  In Section~\ref{cosmosims}, we describe the setup and calibration of our cosmological multifrequency radiation hydrodynamics simulations.  In Section~\ref{results}, we present the results for tracing three different classes of sources and identify which halo mass, stellar metallicity, and age stars are contributing to the hydrogen photoionisation rate as a function of redshift.  In Section~\ref{cavs}, we discuss a few caveats regarding this current generation of simulations.  Finally, in Section~\ref{dc}, we present our discussion and conclusions and describe further uses and anticipated development of the photon tracer algorithm.

\section{A New Algorithm for Tracing Photons in Simulations}
\label{ptalg}

The general question we aim to address is how can we separate the photon contributions of different source populations inside of a numerical simulation and adaptively track the photons emitted by each population as the simulation evolves.  Solving this will allow one to measure the photon contribution of a given source population to a particular  physical process of interest such as reionization or radiative pressure-driven outflows.

\subsection{Algorithm Description}

We focus our current discussion on the implementation in {\small RAMSES-RT} \citep{Teyssier2002,Rosdahl2013} a cosmological, adaptive mesh refinement (AMR) code with coupled radiative transfer (RT); however, our implementation is straightforwardly portable to any other moment-based method.

\subsubsection{Tracing the Photons Emitted by Difference Sources}
For each cell in the simulation and for each of the different multifrequency photon groups, $i$, {\small RAMSES-RT} stores a photon number density, $N_{\gamma,i}$, and a photon flux vector ${\boldsymbol F}_{i}$ that has three directional components (x,y,z).  In order to separate the contributions from different source populations of interest, we create an additional  tracer vector ${\boldsymbol X_i}$ for each of the different multifrequency photon groups that has a dimension equal to the number of different source populations we wish to track in the simulation.  For instance, in the case where we wish to separate the contributions of Pop.~III stars versus Pop.~II stars versus black holes, ${\boldsymbol X_i}=[X_{{\rm Pop. III},i},X_{{\rm Pop. II},i},X_{{\rm BH},i}]$.  This vector holds the fractional abundances of $N_{\gamma,i}$ represented by each of the different source populations.  For example, if a cell contains 100 photons in photon group $i$, 30 of which were contributed by Pop.~III stars, 25 of which were contributed by Pop.~II stars, and 45 of which were contributed by black holes, ${\boldsymbol X_i}=[X_{{\rm Pop. III},i}=0.30,X_{{\rm Pop. II},i}=0.25,X_{{\rm BH},i}=0.45]$.  Crucially, the sum of the elements in ${\boldsymbol X_i}$ must equal 1.

The value of ${\boldsymbol X_i}$ can change only on two\footnote{Here we have used the on-the-spot approximation and thus ignore the contribution from recombination radiation.  This can be accounted for in the cooling routines and added as an additional tracer group.  For this work, we ignore this effect.} occasions:
\begin{enumerate}
    \item during the injection step
    \item during the advection step.
\end{enumerate}
During the injection step, photons emitted by sources are added to the values of $N_{\gamma,i}$ in their host cell.  The values of ${\boldsymbol X_i}$ are updated accordingly to represent the new fractional abundances,
\begin{equation}
{\boldsymbol X_{i,{\rm new}}}=\frac{{\boldsymbol X_i}N_i+{\boldsymbol N_{\rm inj}}}{N_i+\sum_j^{N_*}{\boldsymbol N_{{\rm inj},j}}},
\end{equation}
where ${\boldsymbol N_{\rm inj}}$ is the vector containing the number of newly injected photons from each source and $N_*$ is the number of different source populations we wish to track.  The simulation is initialised with $N_{\gamma,i}=0$ and ${\boldsymbol X_i}=1/N_*$. The initial value of ${\boldsymbol X_i}$ is irrelevant as long as $N_{\gamma,i}=0$ because as soon as photons are injected or advected into the cell, the fractional abundances will be updated accordingly. 

During the advection step, photons move between cells and the photon fluxes across each of the six interfaces are calculated using a Global  Lax-Friedrichs (GLF) intercell flux function.  For the flux across the cell boundary of cells $l$, and $l+1$,
\begin{equation}
\label{GLF}
\mathcal{F}_{{\rm GLF},l+1/2}=\frac{\mathcal{F}_l+\mathcal{F}_{l+1}}{2}-\frac{c_{{\rm sim},l}}{2}(\mathcal{U}_{l+1}-\mathcal{U}_{l}),
\end{equation}
where $\mathcal{U}_l=[N_l,{\bf F}_l]$, $\mathcal{F}_l=[{\bf F}_l,c_{{\rm sim},l}^2\mathbb{P}_l]$, and $c_{{\rm sim},l}$ is the speed of light\footnote{See \protect\cite{Katz2017} for a more detailed explanation of how to deal with a changing speed of light across cell boundaries in the context of the variable speed of light approximation.} used in the simulation for cell $l$. $N_l$ is the number density of photons, ${\bf F}_l$ is the flux vector, and $\mathbb{P}_l$ is the pressure tensor in cell $l$.  In 1D, taking flux moving from left to right to be positive, {\small RAMSES} subtracts the intercell flux at the right face from the intercell flux at the left face to update $N$ and ${\bf F}$ in the cell.  In order to update ${\boldsymbol X_i}$, we must deal with the flux of photons moving into the cell differently from the flux moving out and thus the left face separate from the right face.  If photons move out of the central cell, across one of the six faces, the value of ${\boldsymbol X_i}$ remains unchanged.  If photons move into a cell, the value of ${\boldsymbol X_i}$ is updated assuming that the incoming photons have a source distribution as given by the value of ${\boldsymbol X_i}$ in the adjacent cell.  The following equation summarises the update to ${\boldsymbol X_i}$ for an adjacent cell on the left of the cell of interest,
\begin{equation}
{\boldsymbol X_{i,{\rm new}}}=\frac{{\boldsymbol X_i}N_i+{\boldsymbol X_{i,{\rm adj}}}{\rm max}(\frac{\Delta t}{\Delta x}\mathcal{F}_{{\rm GLF},N},0)}{N_i+{\rm max}(\frac{\Delta t}{\Delta x}\mathcal{F}_{{\rm GLF},N},0)}.
\end{equation}
Here, ${\boldsymbol X_{i,{\rm new}}}$ is the value of ${\boldsymbol X_i}$ after the update, ${\boldsymbol X_{i,{\rm adj}}}$ is the value of ${\boldsymbol X_i}$ in the adjacent cell on the left, and $\frac{\Delta t}{\Delta x}\mathcal{F}_{{\rm GLF},N}$ is the number of photons moving from the cell on the left to the cell on the right (if this quantity is negative photons are moving from the right cell to the left cell).

The code loops over the three physical dimensions and the values of ${\boldsymbol X_i}$ are updated during the loop. The directional splitting scheme in {\small RAMSES} (of storing before- and after-states in each cells) ensures that the tracer advection is completely independent of the order of dimensions (x,y,z) in which the tracers are advected. It further ensures that a tracer element never traverses more than one cell interface in one radiative transfer step. 

Note that no additional calls to the Godunov solver are required for this computation.  This would not be the case if additional photon groups were added to track different source populations.  In terms of computation time, updating ${\boldsymbol X_i}$ is effectively negligible which makes this algorithm very computationally efficient.  For small test problems, including the photon tracers into the computation does not increase the wall-clock time of the simulation by any noticeable amount.  However, the vectors that hold the fractional abundances of photons can add a significant amount to the memory  required ($N_*$ extra variables per photon group).  For large cosmological runs, we have found that the computation can slow down by a modest $20$-$30\%$, which is partially due to the extra communication of variables between cores in the simulations with the parallelised code. 

The photon flux and absorption algorithm is designed to track the number of photons from different sources in each cell and thus the contribution of each source to the instantaneous photoionisation rate.  This is equivalent to measuring which sources are currently responsible for maintaining the ionisation state of a cell.  This is extremely useful for studies of reionization where there exist direct measurements of the photoionisation rate at high redshift \citep{Bolton2007,Wyithe2011,Calverley2011,Becker2013}.  By tracing the photon sources, one can also measure the escape fraction of different source populations within the same galaxy.  Similarly, knowing the number of photons from each source in a given cell also allows the effect of radiation pressure to be split between the different source populations.  This can, for example, be used to compare the relative roles of stellar versus AGN radiation pressure in a given galaxy, which is something we will focus on in upcoming work.

\subsubsection{Measuring Which Source or Process Ionised the Gas}
As described so far, our algorithm does not maintain memory of which photons ionised a cell.  This `history' of the gas is very relevant to probe which sources have actually contributed to reionizing the Universe.  This is different from measuring which sources are contributing to the photoionisation rate.  In order to keep track of this quantity, we have modified the cooling module in {\small RAMSES-RT} to track the history of contributions from different sources to the local ionisation fraction.  For each cell, {\small RAMSES-RT} stores the ionised hydrogen fraction in the variable $x_{\rm HII}$\footnote{When helium is included in the simulation, {\small RAMSES-RT} also stores $x_{\rm HeII}$ and $x_{\rm HeIII}$ but we will only focus on hydrogen due to the multiple ionisation states of helium.}.  This quantity can be split up into its constituent parts such that
\begin{equation}
x_{\rm HII}=x_{\rm HII,CI}+\sum_{i=1}^{N_*}x_{{\rm HII,PI}_i},
\end{equation}
where $x_{\rm HII,CI}$ is the fraction of the gas that was ionised by collisional ionisation, $N_*$ is the number of photon tracer classes, and $x_{{\rm HII,PI}_i}$ is the fraction of gas that was photoionised by group $i$ photons.  The change in the ionisation fraction can then be computed as
\begin{equation}
\label{dxhiidt}
\frac{\partial x_{\rm HII}}{\partial t}=(1-x_{\rm HII})\left[\beta_{\rm HI}n_e+\sum_{j=1}^{N_*}\sum_{i=1}^{N_G}\sigma_{{\rm HI},i}c_{\rm sim}N_{i,j}\right]-x_{\rm HII}\alpha_{\rm HII}n_e,
\end{equation}
where $\beta_{\rm HI}$ is the collisional ionisation rate, $n_e$ is the electron number density, $N_G$ is the number of energy bins for the multifrequency RT, $\sigma_{{\rm HI},i}$ is the cross section of neutral hydrogen for photon energy bin $i$, $c_{\rm sim}$ is the speed of light in the simulation (or the speed of light in a given level if VSLA is used), $N_{i,j}$ is the number density of photons in photon tracer class $j$ and in photon energy bin $i$, and $\alpha_{\rm HII}$ is the hydrogen recombination rate (either case~A or case~B).  The two terms inside the bracket are responsible for the creation of HII while the last term on the right is responsible for the destruction of HII.  Equation~\ref{dxhiidt} is separable so that for collisional ionisation,
\begin{equation}
\frac{\partial x_{\rm HII,CI}}{\partial t}=(1-x_{\rm HII})\beta_{\rm HI}n_e-x_{\rm HII,CI}\alpha_{\rm HII}n_e,
\end{equation}
and for photoionisation from each of the different tracer classes, $i$,
\begin{equation}
\frac{\partial x_{{\rm HII,PI}_i}}{\partial t}=(1-x_{\rm HII})\sum_{j=1}^{N_G}\sigma_{{\rm HI},j}c_{\rm sim}N_{i,j}-x_{{\rm HII,PI}_i}\alpha_{\rm HII}n_e.
\end{equation}
These two equations allow us to track the fraction of the gas that has been ionised by different processes (collisional ionisation and photoionisation) as well as by photons in each of the different photon groups.  Following \cite{Rosdahl2013}, we update the total ionisation fraction in a semi-implicit fashion such that
\begin{equation}
x_{\rm HII}^{t+\Delta t}=x_{\rm HII}^{t}+\frac{\Delta t}{1-J\Delta t}\frac{\partial x_{\rm HII}}{\partial t},
\end{equation}
where $J\equiv\frac{\partial\dot{x}_{\rm HII}}{\partial x_{\rm HII}}$ and $\partial\dot{x}_{\rm HII}$ is the time derivative of the ionised fraction.  This is easily generalised to the collisional ionisation and photoionisation terms so that
\begin{equation}
x_{\rm HII,CI}^{t+\Delta t}=x_{\rm HII,CI}^{t}+\frac{\Delta t}{1-J\Delta t}\frac{\partial x_{\rm HII,CI}}{\partial t},
\end{equation}
and
\begin{equation}
x_{{\rm HII,PI}_i}^{t+\Delta t}=x_{{\rm HII,PI}_i}^{t}+\frac{\Delta t}{1-J\Delta t}\frac{\partial x_{{\rm HII,PI}_i}}{\partial t}.
\end{equation}
We store $x_{{\rm HII,PI}_i}$ as a hydrodynamic variable that is advected between cells in the same way as $x_{{\rm HII}}$.  We do not store $x_{{\rm HII,CI}}$ as this can be calculated from $x_{{\rm HII,PI}_i}$ and $x_{{\rm HII}}$.  This photon absorption algorithm works well in combination with the photon tracer algorithm as we can now measure which sources are driving the photoionisation rate during reionization as well as which sources actually reionized the gas.  The difference is subtle yet important.

Note that we have only applied the photon absorption algorithm to hydrogen.  In principle this can also be done for helium ionisation; however, because HeIII requires two photons to be ionised, keeping track of the different sources responsible for the ionisation becomes more difficult and expensive in terms of memory.  Furthermore, recombinations that form HeII from HeIII become ill defined.  Consider the case where photons from two different sources were absorbed to form HeIII.  If this ion then recombines with an electron to from HeII, there is an ambiguity of which tracer class to assign to the newly formed HeII.  For these reasons, we have only applied the photon absorption algorithm to hydrogen in the current version of the code.  In this manuscript, we will focus only on the photon tracer algorithm and in a follow-up work, we will discuss the results from using the photon absorption algorithm.

\subsubsection{Measuring the Escape Fraction of LyC Photons}
By separating the number density of photons into the various tracer groups, one can measure the globally averaged escape fraction of LyC photons ($\bar{f}_{\rm esc}$) individually for different source populations.  To measure this quantity, we calculate $\bar{f}_{\rm esc}$ at a given time to be 
\begin{equation}
\label{fesc_eqn}
\bar{f}_{\rm esc} = \frac{N_{\gamma,{\rm IGM}}+N_{{\rm absorb,IGM}}}{N_{\gamma,{\rm emitted}}},
\end{equation}
where, $N_{\gamma,{\rm IGM}}$ is the total number of photons present in the IGM, $N_{{\rm absorb,IGM}}$ is the integrated number of photons that were absorbed in the IGM following the formation of the first source down to a given epoch, and $N_{\gamma,{\rm emitted}}$ is the total number of photons emitted by the sources.  This formalism gives the average escape fraction of a given class of sources over a defined time period.  The key to this algorithm is separating gas in galaxies from the IGM in the simulation.  For this, we make the approximation that all gas cells with $\rho_{\rm gas}<180\bar{\rho}_{\rm b}$ are part of the IGM, where $\bar{\rho}_{\rm b}$ is the mean baryonic density as a function of redshift.  In regions affected by SN, the gas density inside of a galaxy can drop below this value and we have added an additional criterion that uses the delayed cooling parameter so that all cells recently affected by SN within the past 20Myr are counted as part of the galaxy and not the IGM, regardless of their density.

The most accurate way to mask galaxies would be to use a halo catalog and check whether a photon resides or an absorption occurs within a cell that is inside the virial radius of a halo.  This would require checking at every time step whether a gas cell belongs to a halo and this is computationally expensive.  There are a few potential issues with using our current method.  There sometimes exists self-shielded gas such as in damped Ly$\alpha$ systems (DLAs) or Lyman limit systems (LLSs) in the IGM that have $\rho_{\rm gas}>180\bar{\rho}_{\rm b}$.  Similarly, a photon may travel through the IGM unimpeded and hit another galaxy.  We expect that as more sources form throughout the course of the simulation, the first effect will be insignificant.  Furthermore, the volume subtended by DLAs and LLSs and other galaxies is significantly smaller than the IGM so we also expect this second effect to be small.

\subsection{Numerical Issues and Pitfalls}
One must take care during the process of refinement (in {\small RAMSES} a cell can be split into $2^{\rm n_{dim}}$ children cells where ${\rm n_{dim}}$ is the number of dimensions in the simulation) and derefinement (in {\small RAMSES} $2^{\rm n_{dim}}$ children cells are merged into a single cell) to ensure that the properties of ${\boldsymbol X_i}$ are conserved accordingly.  Derefinement is rather straightforward and we simply combine the photon number-weighted fractions of each of the children cells to update ${\boldsymbol X_i}$ in the parent cell such that
\begin{equation}
{\boldsymbol X_{i,{\rm parent}}}=\frac{\sum_{i=1}^{2^{\rm n_{dim}}}N_{i,{\rm child}}{\boldsymbol X_{i,{\rm child}}}}{\sum_{i=1}^{2^{\rm n_{dim}}}N_{i,{\rm child}}}.
\end{equation}
However, during the process of refinement, it is common to use a higher order interpolation scheme to account for the gradient that should be present along the newly refined children cells.  Because ${\boldsymbol X_i}$ holds fractional abundances, it is paramount that the vector sums precisely to unity and that all values are greater than or equal to zero.  This is untrue for most interpolation schemes and thus when refining, we prefer a straight injection where all eight children cells hold the same fractional abundance as their parent.  This same method is also used when we perform an advection at a fine-to-coarse boundary.  However, we continue to use a MinMod slope limiter for the other quantities.

A second potentially problematic issue is due to  the use of the M1 closure \citep{Levermore1984} in {\small RAMSES-RT}.  Because photons with fluxes in opposite directions do not cross (see \citealt{Rosdahl2013}), the photon tracers will also suffer the same fate.  It should be emphasised that this is not a weakness of the photon tracer algorithm. When applied to moment-based codes that do not exhibit this feature, there is nothing intrinsic to the photon tracers that prevents mixing.  When interpreting the results of the photon tracers for simulations with the M1 closure, caution must be taken.  However, we expect that globally averaged quantities will nevertheless remain robust and we focus our results on these properties.

For the simulations presented here, we subcycle the radiation step multiple times within a given hydrodynamic time step.  In order for this to be  compatible with the adaptive time stepping used for the AMR grid, one must adopt suitable boundary conditions at the fine-to-coarse interface of AMR levels \citep{Commercon2015}.  For this reason  the number of photons is no longer strictly conserved.  This can affect our tracer algorithm if, for example, the percentage of artificial creations or destructions of photons is different for each different tracer group.  At every time step in the simulation, we can measure the cumulative number of photons emitted by all sources and compare this to the sum of the total number of photons in the simulation and the total number of absorptions that have occurred.  We have checked that when the number of radiation subcycles is restricted to one, the cumulative number of photons emitted matches those that remain in the simulation and that have been absorbed.  For our fiducial simulations presented here, we allow for up to 500 radiation subcycles per hydrodynamic time step (although the simulation often uses far fewer, especially after a supernova has occurred) and can confirm that the use of subcycles only results in an inaccuracy of $\sim6.5\%$ at $z=6$.  This slightly varies between the different tracer groups; however, these deviations are small compared the main features in our results.

Because subcycles no longer conserve photons, we have opted to adjust Equation~\ref{fesc_eqn} so that:
\begin{equation}
\label{fesc_eqn1}
\bar{f}_{\rm esc} = \frac{N_{\gamma,{\rm IGM}}+N_{{\rm absorb,IGM}}}{N_{\gamma,{\rm IGM}}+N_{{\rm absorb,IGM}}+N_{\gamma,{\rm Gal}}+N_{{\rm absorb,Gal}}},
\end{equation}
where $N_{\gamma,{\rm Gal}}$ and $N_{{\rm absorb,Gal}}$ are the total number of photons residing and number of absorptions which have occurred inside galaxies (here defined as cells with $\rho_{\rm gas}\geq 180\bar{\rho}_{\rm b}$).  In the case of only one subcycle, the denominator of Equation~\ref{fesc_eqn1} perfectly matches the cumulative number of photons emitted in the simulation; however, when subcycles are included, the denominator accounts for the artificial creation and destruction of photons and ensures that $\bar{f}_{\rm esc}$ remains between zero and one.

\section{Cosmological Radiation-Hydrodynamics Simulations}
\label{cosmosims}
\subsection{Setup}
The simulations presented in this work follow on from those described in \cite{Katz2017}.  The same set of initial conditions, which have been generated with {\small MUSIC} \citep{Hahn2011}, are used, representing a $10\,{\rm c} {\rm Mpc} \, h^{-1}$ box initialised at $z=150$ with a uniform grid of $256^3$ dark matter particles (the dark matter particle mass is $m_{\rm dm}=6.51\times10^6$M$_{\odot}$).  The gas is assumed to be initially neutral and composed of 76\% hydrogen and 24\% helium by mass.  We have used the following cosmological parameters as given by \cite{Planck2016}: $h=0.6731$, $\Omega_{\rm m}=0.315$, $\Omega_{\Lambda}=0.685$, $\Omega_{\rm b}=0.049$, $\sigma_8=0.829$, and $n_s=0.9655$.  Our fiducial model has been run to $z=4$.

We use the publicly available radiation hydrodynamics code {\small RAMSES-RT} \citep{Rosdahl2013}, which is an extension of the massively parallel cosmological adaptive mesh refinement code {\small RAMSES} \citep{Teyssier2002}.  In this work, we focus on the properties of reionization so we limit the RT to three frequency bins as listed in Table~\ref{ebins} and only use a six-species non-equilibrium chemistry model which tracks H, H$^+$, $e^-$, He, He$^+$, and He$^{++}$.  Star formation and feedback are treated very similarly to \cite{Katz2017}. We use a simple density threshold for star formation and delayed cooling supernova feedback \citep{Teyssier2013} where we shut down cooling for $\sim20$Myr in the cells affected by the SN.  We have made two small modifications compared to \cite{Katz2017}. We increase the density threshold for star formation to $n_{\rm H}=1$cm$^{-1}$ and rather than using the spread feedback model presented in \cite{Katz2017} all feedback energy, mass return, and metals are injected into the host cell of the star particle.  When stars form in the simulations, we inject ionising radiation into their host cells. For this  we adopt the spectral energy distributions (SEDs)  from models including binary stellar evolution (BPASSv2, \citealt{BPASS,Stanway2016}) that assume a maximum stellar mass of 300M$_{\odot}$ and an IMF slope of $-2.35$.  We have introduced an additional free parameter, $f_{\rm lum}$, which represents an unresolved escape fraction, that we use to rescale the ionising luminosities of all stars to reproduce a reionization history consistent with observations (see Section~\ref{gprop}).  The resolution and refinement strategy employed here is exactly the same as presented for the L14-RT model of \cite{Katz2017}, which refines a cell when it contains eight times the mass in either dark matter or baryons that it had at the previous level upon refinement.  The simulation attempts to maintain a constant physical resolution of $\sim125$pc.  We include primordial cooling from collisional ionisations, recombinations, collisional excitation, Bremsstrahlung, Compton cooling (and heating), and dielectronic recombination for H and He and their ions as presented in Appendix~E of \cite{Rosdahl2013}.  For metal line cooling, we use the default cooling model available in {\small RAMSES} that interpolates from cooling tables computed with {\small CLOUDY} \citep{Ferland2013} for temperatures $>10^4$K.  Below this temperature the cooling rates from metals are taken from \cite{Rosen1995} based on \cite{Dalgarno1972}.

\begin{table}
\centering
\begin{tabular}{@{}llll@{}}
\hline
Bin & E$_{\text{min}}$ & E$_{\text{max}}$ & Main Function\\
 & [eV] & [eV] & \\
\hline
1 & 13.60 & 24.59 & H Photoionisation\\
2 & 24.59 & 54.42 & He \& H Photoionisation\\
3 & 54.42 & $\infty$ & He, He$^+$, \& H Photoionisation\\
\hline
\end{tabular}
\caption{Photon energy bins used in the simulation.  E$_{\text{min}}$ and E$_{\text{max}}$ represent the lower and upper energy limits of the photon energy bins used in the simulation.}
\label{ebins}
\end{table}

\begin{figure*}
\centerline{\includegraphics[scale=1.0,trim={0 0.2cm 0 0.6cm},clip]{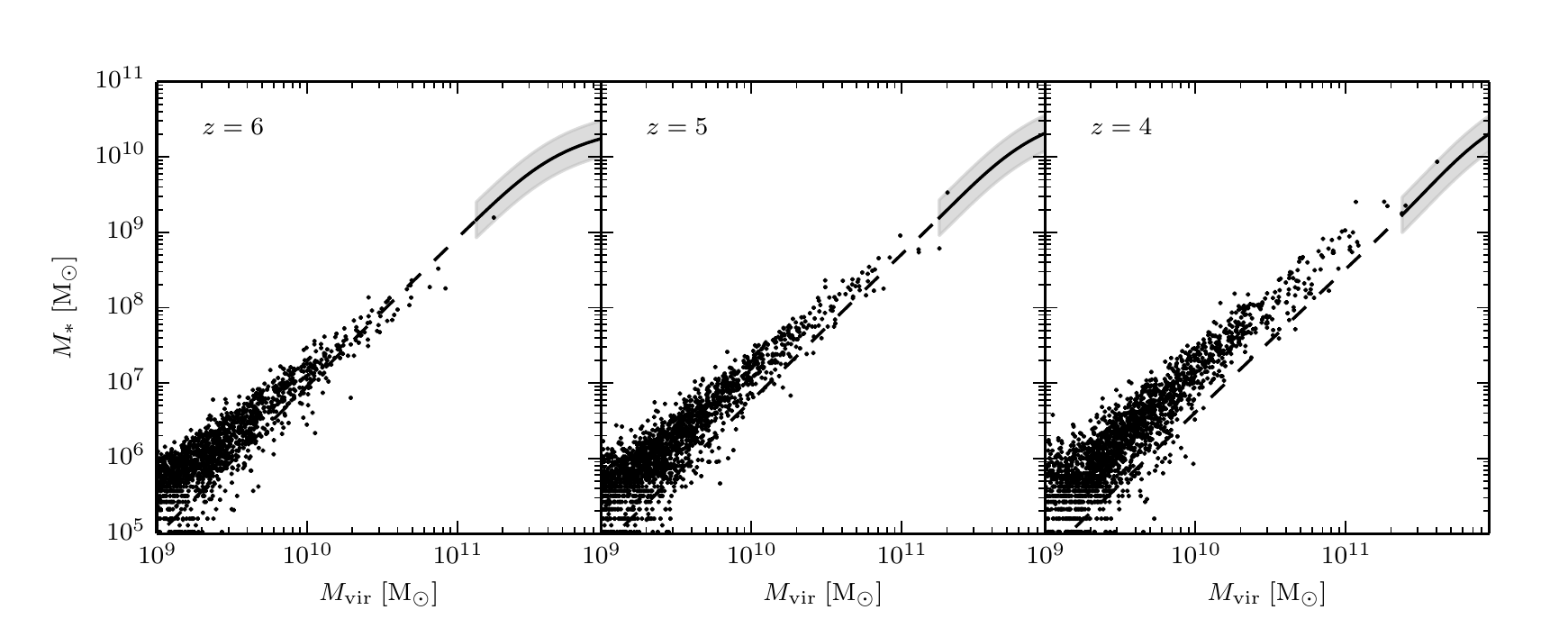}}
\caption{Stellar mass to halo mass relation for galaxies with $M_{\rm vir}>10^9{\rm M_{\odot}}$ at $z=6$ (left), $z=5$ (middle), and $z=4$ (right).  The data points represent haloes in the simulation while the black line and shaded region shows the estimated stellar mass to halo mass relation and 1$\sigma$ scatter as computed by \protect\cite{Behroozi2013}.  The dashed black line shows the extrapolation of the predicted relation from abundance matching.}
\label{SMHM}
\end{figure*}

\cite{Katz2017} presented the variable speed of light approximation (VSLA) to properly model the propagation of ionisation fronts in both low and high density gas.  We have further developed the algorithm for this work so that it is compatible with the RT-subcycling present in {\small RAMSES-RT} \citep{Rosdahl2018}.  This new version of VSLA is significantly faster than the previous version making it more competitive in terms of computational cost for cosmological radiation hydrodynamics simulations.  For the fiducial set of simulations considered here, we set the maximum speed of light to be $0.1c$ ($c=3\times10^8{\rm m\ s^{-1}}$) at the coarse grid and a minimum speed of light of $0.0125c$.  Intermediate refinement levels have a value of $c_{\rm sim}$ interpolated between these two values in factors of two.  Compared to using a standard version of the reduced speed of light approximation (RSLA) with $c_{\rm sim}=0.0125c$ on all refinement levels, our simulations with this method  consume 1.35 times more computing time.  To test the effects of changing the speed of light, we run three additional simulations where we set $c_{\rm sim}=0.0125c$ on all levels, or $c_{\rm sim}=0.2c$ or $c_{\rm sim}=0.4c$ on the coarse grid using the canonical VSLA algorithm.  As we show in Appendix~\ref{SLC}, the photoionisation rates between the VSLA runs are reasonably well converged at $z=6$ for our purposes if  the reionization histories are properly calibrated.  We have thus chosen the  computationally less expensive value of $c_{\rm sim}=0.1c$ on the coarse grid.

\subsection{Calibration}
\label{gprop}
We begin by calibrating our simulations such that they give a reasonable galaxy population as well as reionization history.  The main source of energetic feedback in the simulation is from supernovae (SNe).  In \cite{Katz2017}, we calibrated the SN feedback in order to reproduce the stellar mass-halo mass relation at $z=6$ derived from abundance matching \citep{Behroozi2013} and this is done by changing the time that cooling is shut down in the cells affected by SNe.  We have also slightly altered the star formation algorithm  (by changing the density threshold for star formation) and we use a different cooling function at $T<10^4$K compared to \cite{Katz2017}.  Note, however, that these latter changes have  a minimal effect on star formation and the delayed cooling parameter used here is the same 20Myr that was used in \cite{Katz2017}.  In Figure~\ref{SMHM}, we plot the simulated stellar mass-halo mass relations at $z=6$, $z=5$, and $z=4$ and compare this to the predicted relation and $1\sigma$ scatter from abundance matching \citep{Behroozi2013}.  Note that this relation has been extrapolated to the mass ranges present in our simulation.  Halo catalogs have been created with HOP \citep{Eisenstein1998} and all stars within the virial radius of a halo are assigned to that galaxy.  At both $z=6$ and $z=5$ there is good agreement between our simulations and the estimates from abundance matching, especially for the most massive haloes.  At $z=4$, the stellar masses in our simulations  begin to fall systematically somewhat higher than the relation, although many of the highest mass systems fall well within the $1\sigma$ scatter.  There is a tendency for the slope of our simulated stellar mass-halo mass relation to be slightly shallower than the predictions from abundance matching.  This disagreement seems to go away by $z=4$ where the slopes are  comparable.  Note that our simulation was calibrated  to match the $z=6$ relation so clearly the feedback efficiency is decreasing with decreasing redshift.  The agreement between our simulations and predictions from abundance matching is nevertheless reasonably satisfactory, as the majority of our analysis will be conducted at $z\geq6$.

\begin{figure}
\centerline{\includegraphics[scale=1.0,trim={0 0.8cm 0 1.1cm},clip]{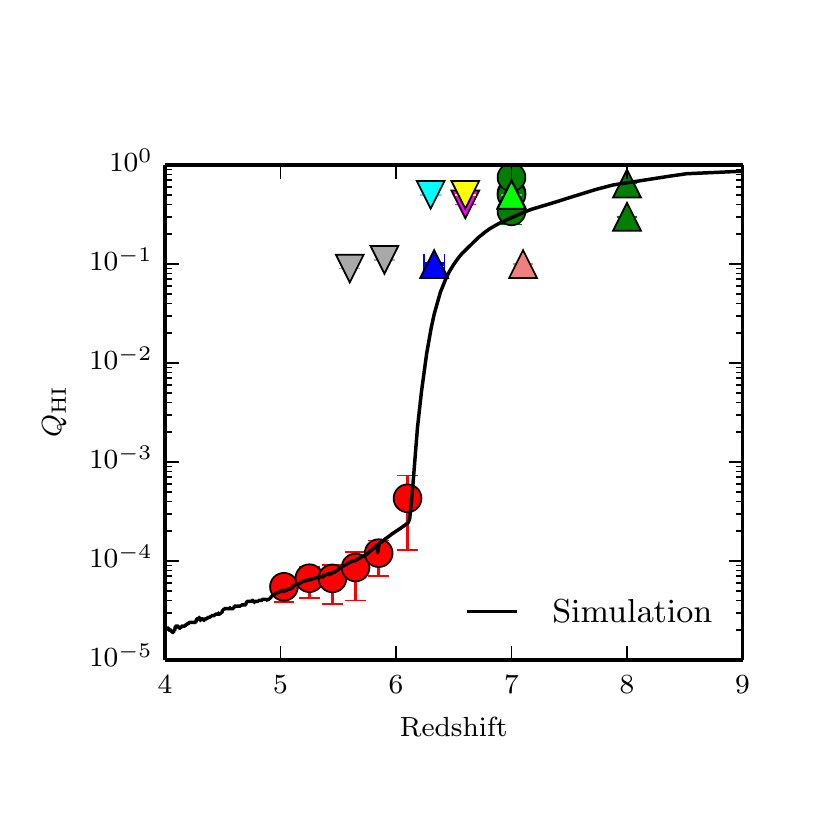}}
\caption{The volume filling factor of neutral hydrogen as a function of redshift in our simulation is shown as the solid black line.  We compare to various (model dependent) measurements of this value compiled by \protect\cite{Bouwens2015}.  Red points are from measurements of the Gunn-Peterson optical depth \protect\citep{Fan2006}, grey is from dark gaps in quasar spectra \protect\citep{McGreer2015}, blue is from Ly$\alpha$ damping wings of quasars \protect\citep{Schroeder2013},  green is from prevalence of Ly$\alpha$ emission in galaxies \protect\citep{Schenker2014,Caruana2014,Ono2012,Pentericci2014,Robertson2013,Tilvi2014}, cyan is from Ly$\alpha$ damping wings of GRBs \protect\citep{Totani2006,McQuinn2008}, magenta is from Ly$\alpha$ emitters \protect\citep{Ouchi2010}, yellow is from galaxy clustering \protect\citep{McQuinn2007,Ouchi2010}, orange is from Ly$\alpha$ emitter luminosity functions \protect\citep{Ota2008}, lime is from clustering of Ly$\alpha$ emitting galaxies \protect\citep{Sobacchi2015}, and light coral is from quasar near-zones \protect\citep{Mortlock2011,Bolton2011}.  Downward triangles are upper limits while upward triangles are lower limits.  Our simulation reproduces the observations in both the timing of reionization as well as the ionised fraction in the post-reionization Universe reasonably well.}
\label{xHIz4}
\end{figure}

Although our choice of SED defines the luminosity of our star particles, due to finite spatial and mass resolution and limited box size, we cannot resolve all of the sources that contribute to reionization, nor do we resolve the small-scale ISM physics that governs the escape fraction of Lyman continuum (LyC) photons.  For this reason, we have introduced a free parameter, $f_{\rm lum}$, that modulates the luminosity of the star particles and can be tuned to calibrate the reionization history (see e.g., \citealt{Gnedin2014,Pawlik2017}).  For our fiducial model, we have set  $f_{\rm lum}=1.193$.  In Figure~\ref{xHIz4}, we plot the volume filling factor of neutral hydrogen as a function of redshift, down to $z=4$.  We compare this to observations of the volume filling fraction of neutral hydrogen derived from measurements of the Gunn-Peterson optical depth \citep{Fan2006}, dark gaps in quasar spectra \citep{McGreer2015}, Ly$\alpha$ damping wings of quasars \citep{Schroeder2013}, prevalence of Ly$\alpha$ emission in galaxies \citep{Schenker2014,Caruana2014,Ono2012,Pentericci2014,Robertson2013,Tilvi2014}, Ly$\alpha$ damping wings of GRBs \citep{Totani2006,McQuinn2008}, Ly$\alpha$ emitters \citep{Ouchi2010}, galaxy clustering \citep{McQuinn2007,Ouchi2010}, Ly$\alpha$ emitter luminosity functions \citep{Ota2008}, clustering of Ly$\alpha$ emitting galaxies \citep{Sobacchi2015} and quasar near-zones \citep{Mortlock2011,Bolton2011} as compiled by \cite{Bouwens2015}.  The agreement between our simulation and the observational constraints is very good, especially compared to the tightest constraints in the post-reionization epoch from measurements of the Gunn-Peterson optical depth \citep{Fan2006}.  It should be emphasised that this calibration is far from trivial as the reionization history is extremely sensitive to the choice of $f_{\rm lum}$ (see e.g., \citealt{Gnedin2014,Pawlik2017}).  The value of $f_{\rm lum}$ is slightly dependent on the chosen value of the speed of light, in particular after the volume filling factor has reached 50\%. This is further discussed in Appendix~\ref{SLC}.  

Because $f_{\rm lum}$ is greater than one, our simulation requires more photons than assumed from the stellar evolution model in order to result in a reasonable reionization history (see also \citealt{Chardin2015}).  The lowest mass atomic-cooling haloes, with $M_{\rm halo} \sim  10^8 \, {\rm M}_\odot$, are not resolved by our simulation and these lower mass haloes may indeed be important for reionization \citep[e.g.,][]{Kimm2017}.  Hence it might be unsurprising that $f_{\rm lum}>1$.  Similarly, with finite spatial resolution, we do not resolve the multiphase structure in the ISM of our galaxies.  Thus, the highly ionised channels leading out of the galaxy that might be present in higher resolution simulations are not resolved by our simulation and this may artificially decrease the escape fraction of LyC photons.  Limited spatial resolution could also work in the opposite way in that we do not resolve the highest density regions of the ISM that may act as sinks for LyC photons which could indeed increase the effective $f_{\rm esc}$ in our simulation.  Finally, because of finite box size, our simulation does not capture the most massive haloes that are present in the Universe at $z=6$.  If these most massive systems play an important role in reionization, our simulation will not be able to capture their emission.  Thus $f_{\rm lum}$ would need to be increased in order to account for these missing photons.  Clearly, the value of $f_{\rm lum}$ is a convolution of a number of different numerical limitations of our simulation and this is indeed a caveat of the work presented here (see \citealt{Rosdahl2018} for further discussion of this point).

\begin{figure}
\centerline{\includegraphics[scale=1.0,trim={0 0.8cm 0 1.1cm},clip]{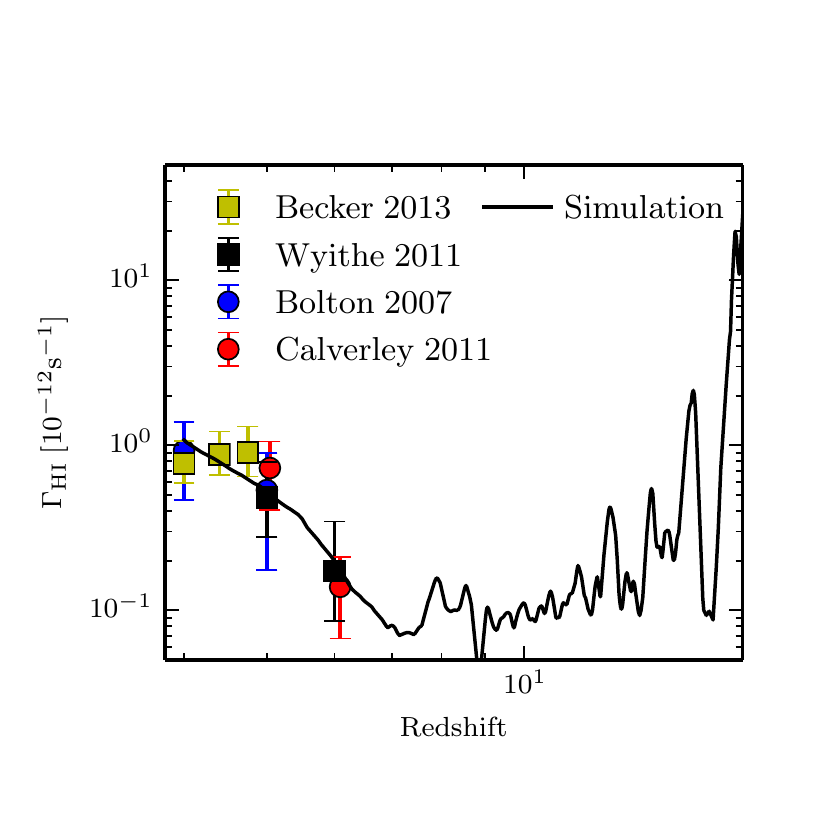}}
\caption{The volume-weighted hydrogen photoionisation rate in ionised regions as a function of redshift is shown as the solid black line.  We define an ionised region as those cells that are at least 50\% ionised.  Observational constraints are shown as the data points and have been adopted from \protect\cite{Bolton2007,Wyithe2011,Calverley2011,Becker2013}.  Our simulation reproduces the observed value of $\Gamma_{\rm HI}$ reasonably well.  At $z<5$ the slope of the evolution  $\Gamma_{\rm HI}$ in our simulation is somewhat steeper than observed. Note that our simulations do not include any contribution to the ionising emissivity from QSOs which should increasingly contribute towards lower redshift.}
\label{gammaz4}
\end{figure}

Note again that we have calibrated the ionising luminosities in our simulations such that   the volume-weighted mean photoionisation rate of neutral hydrogen in ionised regions, $\Gamma_{\rm HI}$  matches observations\footnote{This is calculated for an individual cell as $N_{\gamma}c_{\rm sim}(l)\sigma_{\rm HI}$ where $N_{\gamma}$ is the number density of ionising photons, $c_{\rm sim}(l)$ is the speed of light in the cell, and $\sigma_{\rm HI}$ is the cross section of neutral hydrogen.  We sum over all of the photon energy bins and perform a volume-weighted average across the simulation volume.}.  Our photon tracer algorithm is designed to measure the contribution of each different type of source to this value and hence, agreement with observations is very important.  In Figure~\ref{gammaz4}, we plot $\Gamma_{\rm HI}$ as a function of redshift down to $z=4$ measured directly from our simulation and compare this with observations of $\Gamma_{\rm HI}$ measured from quasar spectra \citep{Bolton2007,Wyithe2011,Calverley2011,Becker2013}.  Once again, we find good agreement between our simulation and the observed data at $z\geq5$.  At the highest redshifts shown in Figure~\ref{gammaz4}, $\Gamma_{\rm HI}$ is often very high (i.e., $>10^{12}$s$^{-1}$) and fluctuates.  Very little of the volume in the simulation is ionised and thus $\Gamma_{\rm HI}$ is probing regions very close to individual galaxies.  As the ionisation fronts begin to move away from the sources, $\Gamma_{\rm HI}$ decreases as the photons are spread over a larger volume and becomes smoother as more star formation occurs.

\begin{table*}
\centering
\begin{tabular}{@{}llll@{}}
\hline
Tracer Property & T1 & T2 & T3\\
Colour in Plots & {\textcolor{red}{Red}} &  {\textcolor{green}{Green}} &  {\textcolor{blue}{Blue}} \\
\hline
Halo Mass & $M\leq10^9{\rm M_{\odot}}h^{-1}$ & $10^9{\rm M_{\odot}}h^{-1}<M\leq10^{10}{\rm M_{\odot}}h^{-1}$ & $M>10^{10}{\rm M_{\odot}}h^{-1}$ \\
Stelar Metallicity & $Z\leq10^{-3}Z_{\odot}$ & $10^{-3}Z_{\odot}<Z\leq10^{-1.5}Z_{\odot}$  & $Z>10^{-1.5}Z_{\odot}$\\
Stellar Age & Age $\leq$ 3Myr & 3Myr $<$ Age $\leq$ 10Myr & Age $>$ 10Myr\\
\hline
\end{tabular}
\caption{Photon tracer bins used in the three different simulations.  The most massive halo has mass $M{\rm=1.2\times10^{11}M_{\odot}}$ and the most metal enriched star has metallicity $Z=10^{-1.2}Z_{\odot}$.}
\label{photbins}
\end{table*}

At $z\lesssim7.5$, $\Gamma_{\rm HI}$ begins to increase again.  By this point in the simulation, the HII bubbles have begun to merge and the volume filling ionisation fraction reaches $\sim50\%$, hence $\Gamma_{\rm HI}$ increases as the mean free path for ionising photons rises rapidly.  This continues well into the post-reionization epoch.  Overall the photoionisation rates in our simulation are in reasonable agreement with measured photoionisation rates at $z<6$,  even though the slope of the evolution $\Gamma_{\rm HI}$ is somewhat steeper  than observed at $z\lesssim5$.  At this stage, the evolution of $\Gamma_{\rm HI}$ is observed to be flat while the $\Gamma_{\rm HI}$ in our simulations tends to still increase.  Our analysis of the stellar mass-halo mass relation showed that we likely produce slightly too many stars between $z=6$ and $z=4$ since our simulations match the predictions from abundance matching at $z=6$ but marginally over predict the expected stellar masses at $z=4$.  This could be the reason that $\Gamma_{\rm HI}$ increases more strongly during this epoch than expected from observations.  Reproducing a flat $\Gamma_{\rm HI}$ during this epoch is challenging and may require a contribution from QSOs which are not modelled here (see e.g., \citealt{Chardin2015}).  We should also note here that the slope of the evolution of $\Gamma_{\rm HI}$ in the post-reionization epoch  depends on the assumed speed of light in the simulation, but  as already discussed, this is not the case before the percolation of HII regions at $z>6$ where the focus of our analysis lies.  After percolation, the simulation box is nearly completely ionised and the mean free path of ionising photons increases rapidly. Reducing the speed of light therefore limits the rate of change of the photoionisation rate in this regime.  For reasons of computational cost, for our fiducial simulation we have chosen to use $c_{\rm sim}=0.1c$ on the base grid and thus the slope we measure for $\Gamma_{\rm HI}$ at $z<6$ is shallower than what we would  get if we use a higher  value of the speed of light (see Appendix~\ref{SLC}).  However, we show in Appendix~\ref{SLC} that for different values of $c_{\rm sim}$, as long as the reionization histories are reasonably well converged, so is the volume-weighted mean photoionisation rate in the ionised regions at $z=6$.  This suggests that our analysis during the reionization epoch is rather insensitive to a reasonable choice of $c_{\rm sim}$ and thus this is where we focus the majority of our analysis.  Only during and after percolation does our choice for the speed of light limit what we can infer from the photoionisation rates in our simulation.  In the next Section, we will use our  photon tracer algorithm to investigate how different source populations (i.e., halo mass, stellar metallicity, and stellar population age) contribute to $\Gamma_{\rm HI}$ as a function of redshift in order to better understand how the meta-galactic UV background develops during reionization.

\begin{figure}
\centerline{\includegraphics[scale=1.0,trim={0 0.8cm 0 0.8cm},clip]{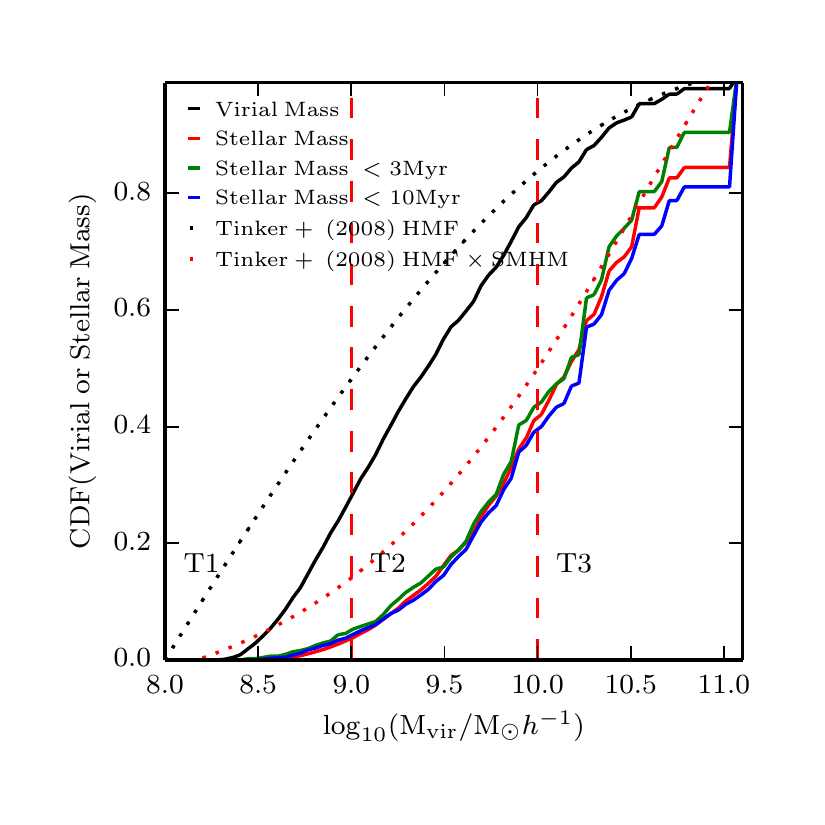}}
\caption{The cumulative distribution function (CDF) of the total mass in dark matter haloes is shown in black while theoretical prediction from \protect\cite{Tinker2008} is shown as the black dotted line.  Shown in red is the CDF of the total stellar mass contained in the halo masses given on the x-axis.  The dotted red line represents the expectation when combining a theoretical halo mass function (HMF) with the $z=6$ SMHM from our simulation.  The green and blue lines represent CDFs of stars with ages $<3$Myr and $<10$Myr, respectively.  These populations represent the stars that emit the majority of ionising photons.  The vertical dashed red lines represent the intervals for the three tracer classes T1, T2, and T3.  Due to resolution, we under predict the total mass in low mass haloes and since these are the least resolved galaxies, we potentially under-predict their contribution to reionization by the difference between the solid and dotted red lines.}
\label{massfunc}
\end{figure}

\section{Results}
\label{results}

In this section, we demonstrate how the contribution to the photoionisation rate can be measured for each individual source population in the simulation with our new photon tracer algorithm.  The quantities we trace and the corresponding  photon tracer bins are listed in Table~\ref{photbins} along with the colours that represent each bin in all of the plots.

\begin{figure*}
\centerline{\includegraphics[scale=1]{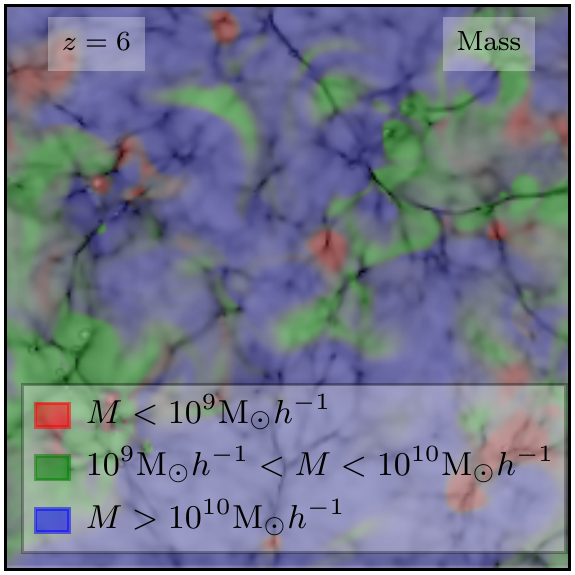}\includegraphics[scale=1]{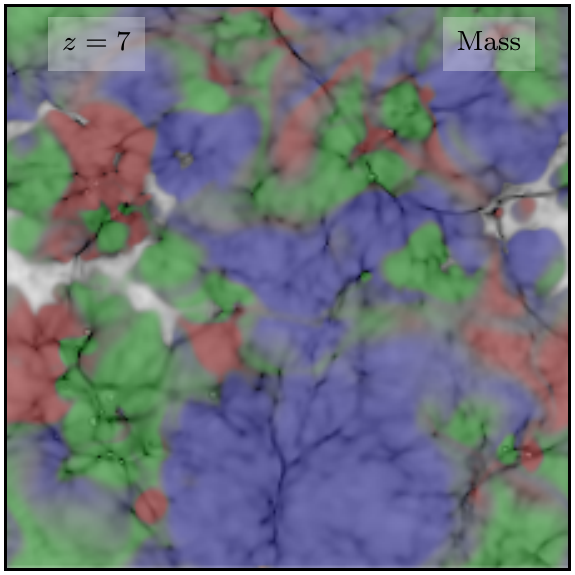}\includegraphics[scale=1]{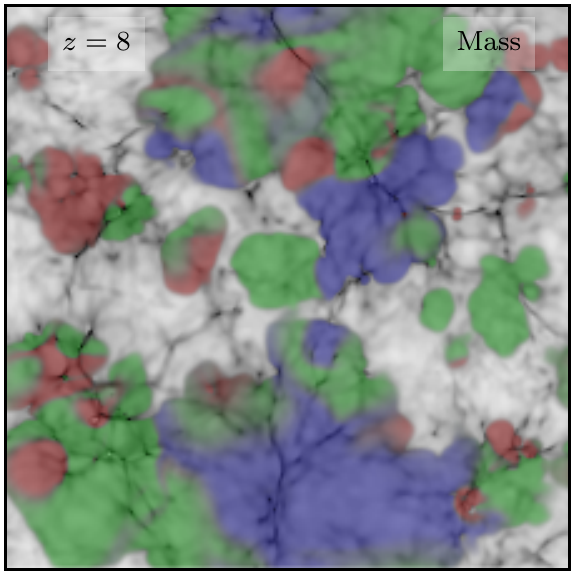}}
\centerline{\includegraphics[scale=1]{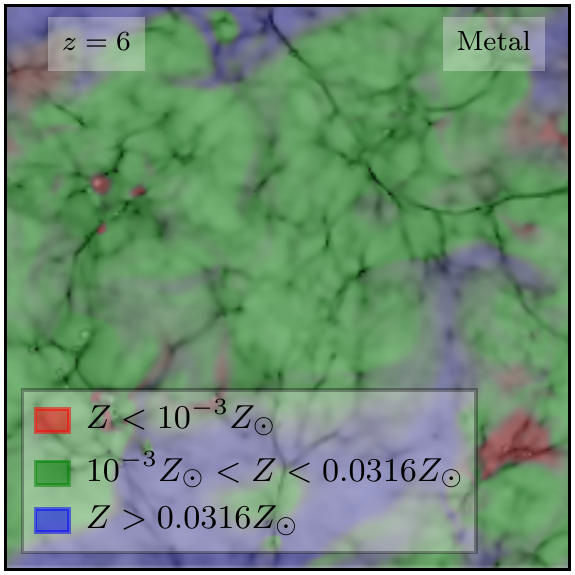}\includegraphics[scale=1]{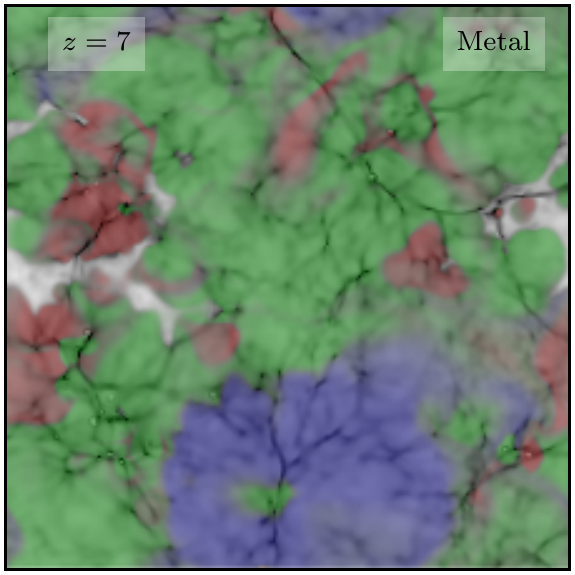}\includegraphics[scale=1.0]{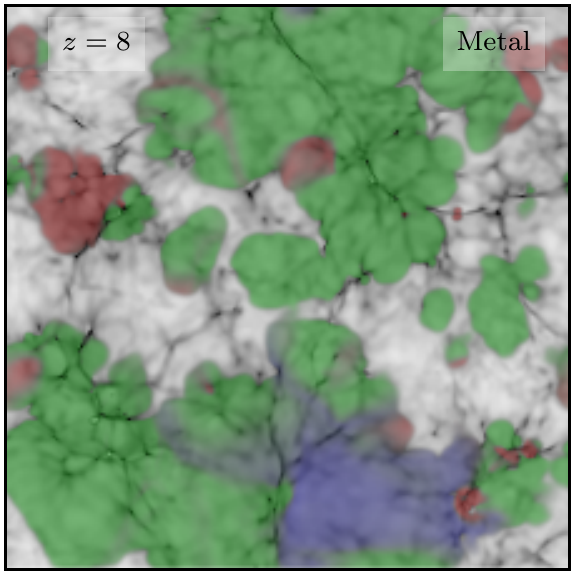}}
\centerline{\includegraphics[scale=1]{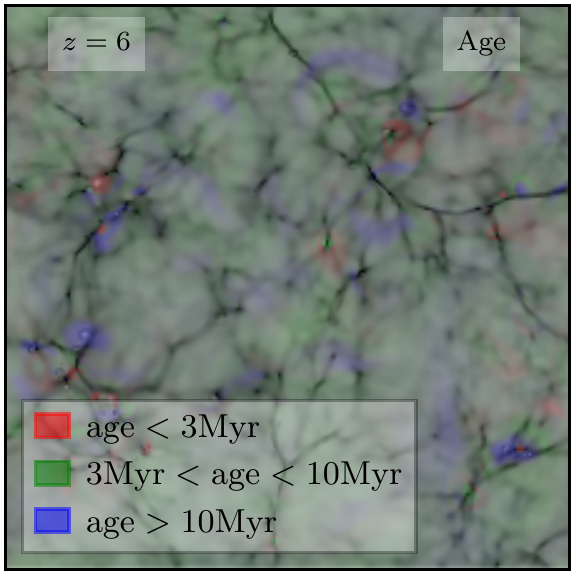}\includegraphics[scale=1]{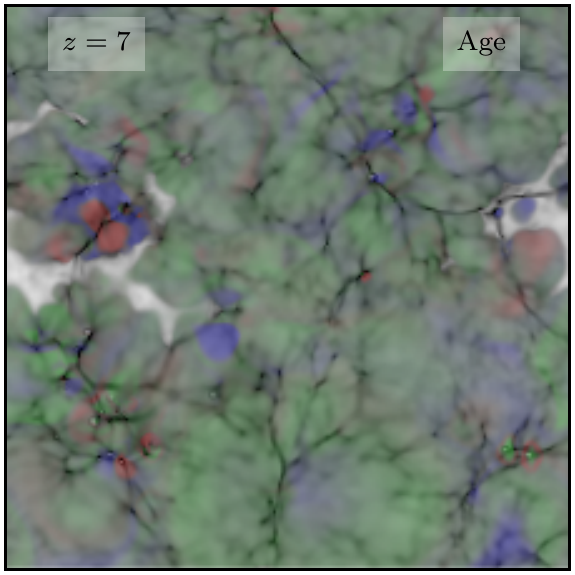}\includegraphics[scale=1]{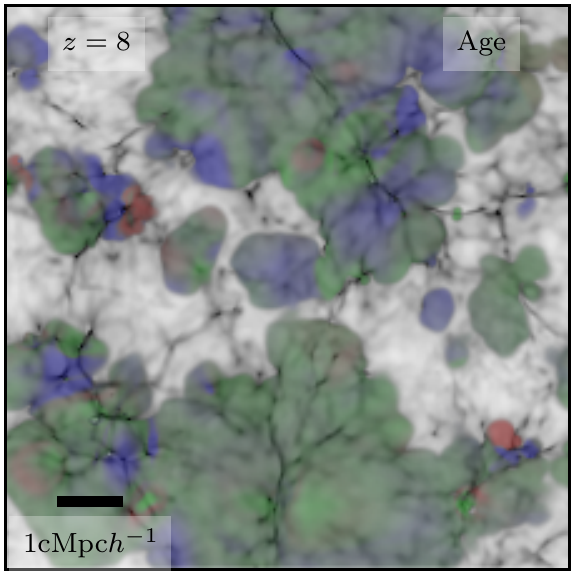}}
\caption{Ionising photon maps coloured by the different tracer classes for a thin slice at the centre of the simulation box.  Coloured regions represent the HII fraction multiplied by the tracer fraction.  This illuminates which tracer class photons are filling the ionised bubble.  The top row shows the run where we trace photons based on halo mass, the middle row shows the tracers based on stellar metallicity, and the bottom row shows the tracers based on stellar age.  In all three rows, red, green, and blue represent photons in bins T1, T2, and T3, respectively as defined in Table~\ref{photbins}.  The underlying gas density is shown in greyscale.  The colours are the most well mixed in the bottom row as the age tracers are not particularly biased to any individual sources.  In contrast, the top two rows are less well mixed and the effects of M1 can be seen.   Averaging over the entire simulation volume should nevertheless give an unbiased result for the contribution of each tracer to the photoionisation rate.}
\label{z8mm}
\end{figure*}

\subsection{Tracing Mass}
One of the most interesting and important questions regarding a galaxy dominated reionization model is which mass galaxies are responsible for providing the bulk of the photons that drive reionization.  To address this question, we apply halo mass-dependent photon tracers to the simulation to track  how galaxies of different mass are contributing to the global photoionisation rate of neutral hydrogen.  Using halo catalogs created with HOP \citep{Eisenstein1998} from dark matter simulation outputs generated at $\Delta z=0.2$ at $z>10$ and $\Delta z=0.05$ at $z<10$, we rerun the simulation with the halo mass-based photon tracers.  We assign star particles to the nearest halo subject to the condition that $r_*\leq r_{\rm vir}$ (where $r_*$ is the distance of the star to the centre of a halo and $r_{\rm vir}$ is the virial radius of the halo) which ensures that subhaloes are taken into account.  A star particle is assigned to a halo both when it forms, as well as when the halo catalog is updated in the simulation which accounts for both halo growth and mergers.  

\begin{figure*}
\centerline{\includegraphics[scale=1.0,trim={0 1.5cm 0 1cm},clip]{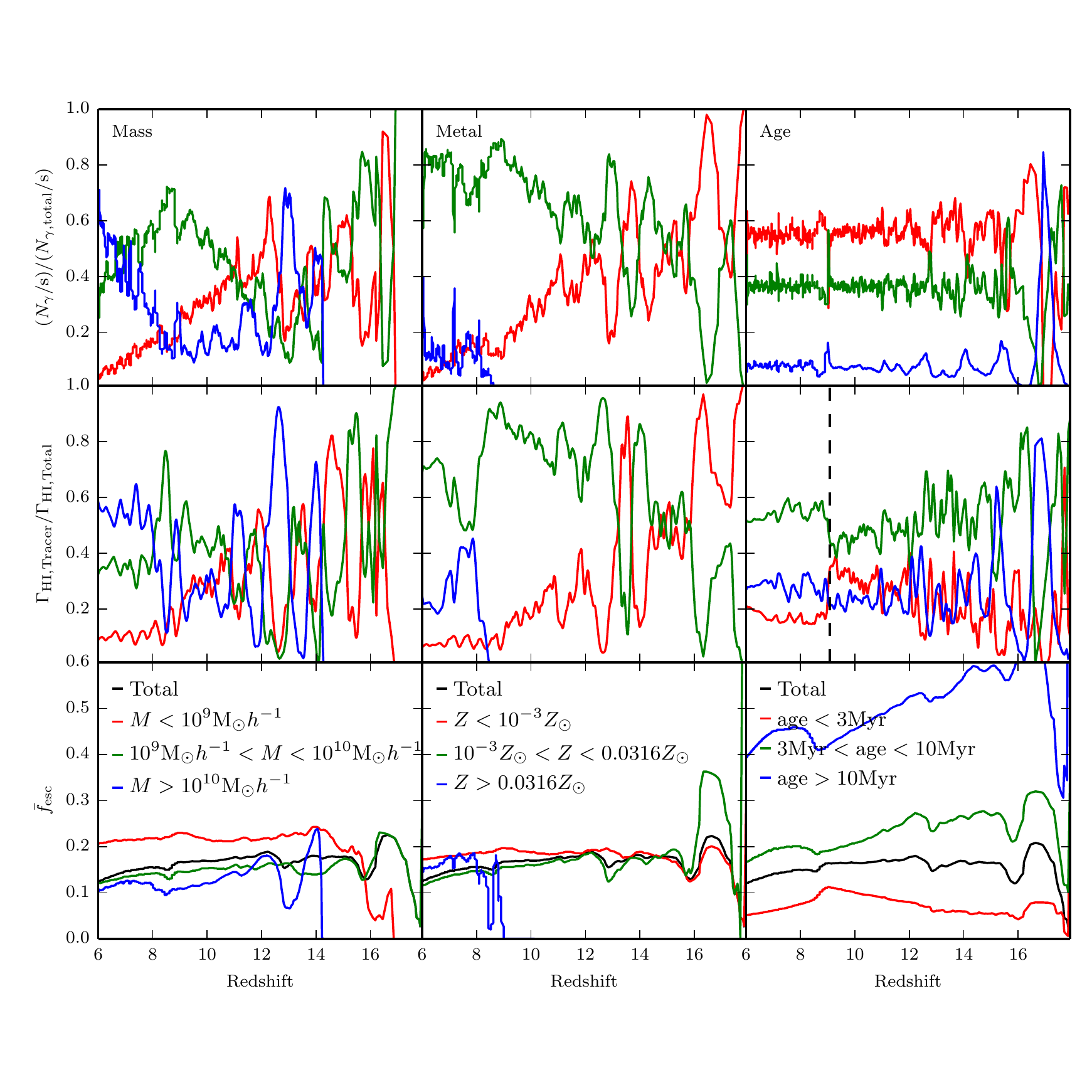}}
\caption{{\it Top}. Fractional contribution of each of the different photon tracers to the instantaneous luminosity as a function of redshift.  The left, centre, and right panels show the results for the halo mass tracers, stellar metallicity tracers, and stellar age tracers, respectively.  The red, green, and blue lines represent sources in T1, T2, and T3, respectively. {\it Centre}. Fractional contribution of each of the different photon tracers to the volume-weighted hydrogen photoionisation rate in ionised regions (defined where $x_{\rm HII}>0.5$) as a function of redshift.  {\it Bottom}. Luminosity-weighted average escape fractions as a function of redshift for each of the different photon tracers.  The black line represents the luminosity-weighted average of all sources in the simulation.  This is the same for all three simulations.}
\label{combined_stats}
\end{figure*}

In Figure~\ref{massfunc}, we show the  cumulative distribution function (CDF) of the total mass in haloes and mass in stars hosted by those haloes split by stellar age at $z=6$.  Although the highest mass tracer group only contains $\sim20\%$ of the total halo mass, it hosts $\sim60\%$ of the total mass in stars in the simulation.  Note that the lowest mass haloes in the simulation are the least resolved so we will under-predict both their number and total stellar content.  The dotted black line in Figure~\ref{massfunc} is an estimate of the CDF of the total mass in haloes from the mass function of \cite{Tinker2008} and the dotted red line shows the expected stellar mass when we apply the stellar mass-halo mass function derived from our simulations at $z=6$ to the \cite{Tinker2008} halo mass function.  This gives an indication of how many photons we may be missing from the lowest mass galaxies at $z=6$.

In the top row of Figure~\ref{z8mm}, we show a thin slice through the centre of the box at $z=6$, $z=7$, and $z=8$ and highlight in different colours the ionised regions where the photoionisation rate is dominated by the different halo mass tracer classes.  By examining the properties of the different regions, we can begin to better understand how the UV background is formed.  At $z=8$, much of the simulation volume  is still  neutral and therefore much of the map  is grey, indicating the presence of mostly neutral gas.  By counting the number of independent HII bubbles of each colour, we can gain insight into the different properties of the sources.  For example, in the $z=8$ slice, there are $\sim12$ unconnected red regions (representing low mass haloes) while there are far fewer green or blue HII bubbles (representing intermediate and high mass haloes, respectively).  The same is the case  for the other two redshift slices as well.  More generally, the red HII bubbles tend to be a lot smaller than the green or blue bubbles.  There are significantly more low mass haloes than intermediate or high mass haloes and likewise, the low mass haloes are less luminous which explains the observed morphology.  However, these HII bubbles do not necessarily share the same photoionisation rate.  By $z=6$, most of the slice is filled with photons that have been emitted by the highest mass haloes and this is indeed representative of the entire box (see Section~\ref{gammavf}).   

\begin{figure*}
\centerline{\includegraphics[scale=1.0,trim={0 0.6cm 0 0.6cm},clip]{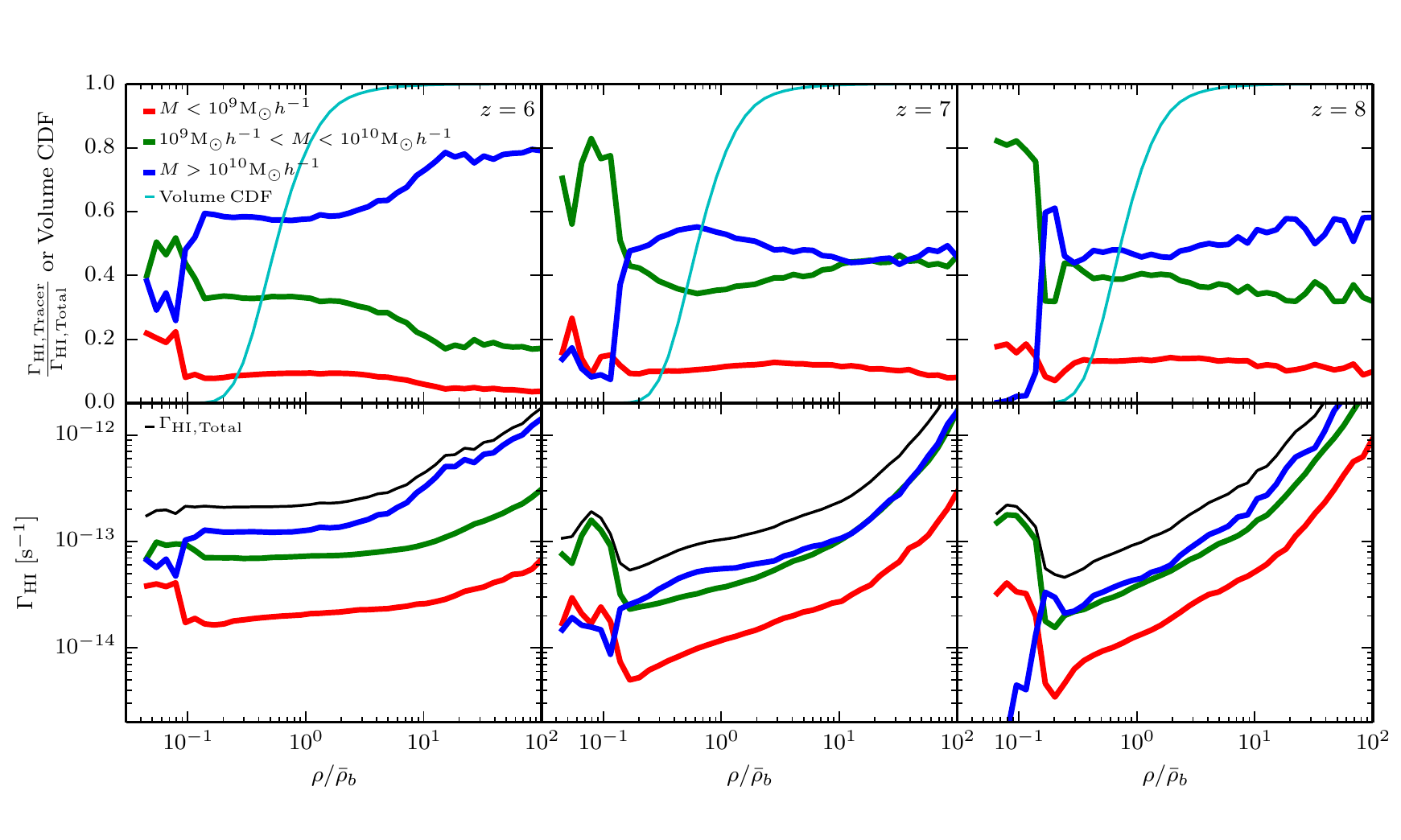}}
\caption{The red, green, and blue lines represent the fractional contribution of each of the different mass tracers to the photoionisation rate in ionised regions as a function of over-density (top) or the absolute photoionisation rate in ionised regions as a function of over-density (bottom).  The black lines in the bottom panels represent the volume-weighted mean photoionisation rate in the ionised regions (defined where $x_{\rm HII}>0.5$).  These values were computed in 50 log-spaced bins at $0.01<\rho/\bar{\rho}_{b}<100$.  Each bin includes all gas cells which have a density in the range $\pm2\%$ of the centre of the bin.  Bins that contain fewer than 100 cells are discarded.  The cyan lines represent the CDF of the total volume of the simulation contained in cells with density less than a certain value.  The majority of the volume of the simulation has $0.1<\rho/\bar{\rho}_{b}<10$.  The photoionisation rate is enhanced in both the lowest density gas (formed via supernova) and the highest density gas (near galaxies).  The left, centre, and right panels show the results at $z=6$, $z=7$, and $z=8$, respectively.}
\label{gamrho_mass}
\end{figure*}

The colours often map out  well defined regions in the box, even after the HII bubbles have overlapped.  This suggests that the photons from different classes of sources are not mixing particularly well.  This is a direct result of using the M1 closure in an optically thin regime.  In the following two subsections, we present simulations where multiple different classes of sources reside in the same HII bubble and hence the mixing improves drastically.  Note that when averaging over the entire volume of the simulated domain to measure the photoionisation rate, it is irrelevant whether the photons have mixed. 

In order to understand how each halo mass bin contributes to the metagalactic UV background as a function of redshift, we have investigated how many photons are emitted by each group.  In the top left panel of Figure~\ref{combined_stats}, we plot the fractional luminosity (i.e. the  luminosity of each tracer group divided by the total) as a function of redshift.  At the highest redshifts ($z>14$), the total luminosity is dominated by photons emitted by the lowest mass and intermediate mass systems.  Because there are only a few star forming haloes in the box, the quantities fluctuate as stars are formed in different mass haloes at different times.  At $z\sim14$, one halo has become massive enough to  fall into the highest mass bin and thus the blue line increase at this redshift.  The contribution from the lowest mass sources decreases steadily from $z=12$ to $z=6$ as haloes become more massive and dwarf galaxy formation is suppressed due to stellar feedback and reionization.  By $z=6$, the highest mass and intermediate mass systems have cumulatively emitted the same number of photons (see Appendix~\ref{luminfo}) while the lowest mass systems have emitted far fewer.  However, the intermediate mass systems emit their photons at an earlier epoch and thus the highest mass systems dominate the luminosity at $z=6$.  In the 10Myr period leading up to $z=6$, the highest mass systems have an average contribution of $\sim64\%$ to the total luminosity.

Most important for measuring the contribution of each halo mass bin to the UV background is determining what fraction of the emitted photons actually escape the galaxy (see Equation~\ref{fesc_eqn1}).  In the bottom left panel of Figure~\ref{combined_stats}, we plot the escape fraction as a function of redshift for the different halo mass tracer groups.  We find that at $z=6$, the luminosity weighted escape fractions of all photons emitted by low, intermediate, and high mass systems are 20.7\%, 12.1\%, and 10.8\%, respectively.  A decreasing escape fraction with increasing halo mass is consistent with various other works that have simulated these systems at much higher spatial resolution \citep{Wise2014,Xu2016,Kimm2017}.  This implies that although the lowest mass systems have the lowest luminosity at $z=6$, their fractional contribution to the UV background will be greater than their fractional luminosity suggests because of their higher escape fraction.

In the centre panel of the left column of Figure~\ref{combined_stats}, we show the fractional contribution to the volume-weighted hydrogen photoionisation rate as a function of redshift for each halo mass bin.  The lowest mass systems maximise their contribution at $z\sim14.5$ when they contribute $\sim85\%$ of the photons to the UV background.  On average, the contribution from the lowest mass systems decreases with time as more stars are formed within more massive haloes.  By $z=6$ the lowest mass haloes represent $\sim10\%$ of the total photoionisation rate in the simulation.  This value is more than twice as large as their average contribution to the total luminosity in the 10Myr preceding $z=6$ and is a result of a substantially higher escape fraction compared to the intermediate and high mass haloes.  At $z=6$, the intermediate mass systems contribute $\sim32\%$ of the total photoionisation rate while the highest mass systems contribute $\sim58\%$.

In addition to the global contribution of each mass tracer bin to the photoionisation rate, it is instructive to understand how each mass bin is contributing to the photoionisation rate at a given gas density.  This is shown in the top row of Figure~\ref{gamrho_mass} in terms of the fractional contributions and in the bottom row in terms of the absolute photoionisation rate.  Beginning with the lowest density gas ($\rho/\bar{\rho}_{\rm b}<0.1$), the photoionisation rate in these regions is enhanced above the value at the mean density (see the bottom row of Figure~\ref{gamrho_mass} where this is most pronounced at $z=7$ and $z=8$) and is dominated by the intermediate mass systems at all three redshifts.  This gas is metal rich and we find that much of it is in the vicinity of recent SN.  The SN are effective in driving the gas to a lower density and maintaining that state in the lower mass haloes, hence the lower mass systems have an enhanced contribution to the photoionisation rate at these densities.  

Focusing on gas at intermediate densities ($0.1<\rho/\bar{\rho}_{\rm b}<10$), at $z=6$, the relative contribution to the photoionisation rate from each tracer bin is constant.  This density range covers nearly the entire volume as can be seen by the steep volume CDF plotted as the cyan lines in the top row of Figure~\ref{gamrho_mass}.  This shows the cumulative fraction of the simulation volume that exists at densities below that given on the x-axis.  At $z=6$, the total photoionisation rate is also nearly constant in this density interval (see the bottom left panel of Figure~\ref{gamrho_mass}).  The box is completely ionised and thus a constant photoionisation background has formed independent of density in this regime.  At $z=7$, the photoionisation rate in the ionised regions increases by a factor of $\sim3$ between $\rho/\bar{\rho}_{\rm b}=0.1$ and $\rho/\bar{\rho}_{\rm b}=10$ because the box is not fully ionised. 

The highest density gas ($\rho/\bar{\rho}_{\rm b}>10$) is probing ionised regions in and around galaxies.  Here, the photoionisation rate is enhanced compared to the lower density gas (see also \citealt{Chardin2017b}). At $z=6$, in this regime, the fractional contribution from the highest mass systems increases while contributions from the intermediate mass and lowest mass systems decrease as a function of density at $\rho/\bar{\rho}_{\rm b}>10$.  This highest mass systems are the most luminous and therefore we expect these to dominate $\Gamma_{\rm HI}$ at high densities.

\subsection{Tracing Metallicity}
A valuable property of the photon tracer algorithm is that one can use it to study different source populations that reside in the same halo.  Measuring this analytically is extremely difficult because one would have to make an additional assumption for the escape fraction of a given source population for each type of halo as a function of redshift and this is unconstrained observationally.  To exploit this property of the photon tracers, we now group the photons into different tracer bins based on the metallicity of the stellar population that emitted them.  

\begin{figure}
\centerline{\includegraphics[scale=1.0]{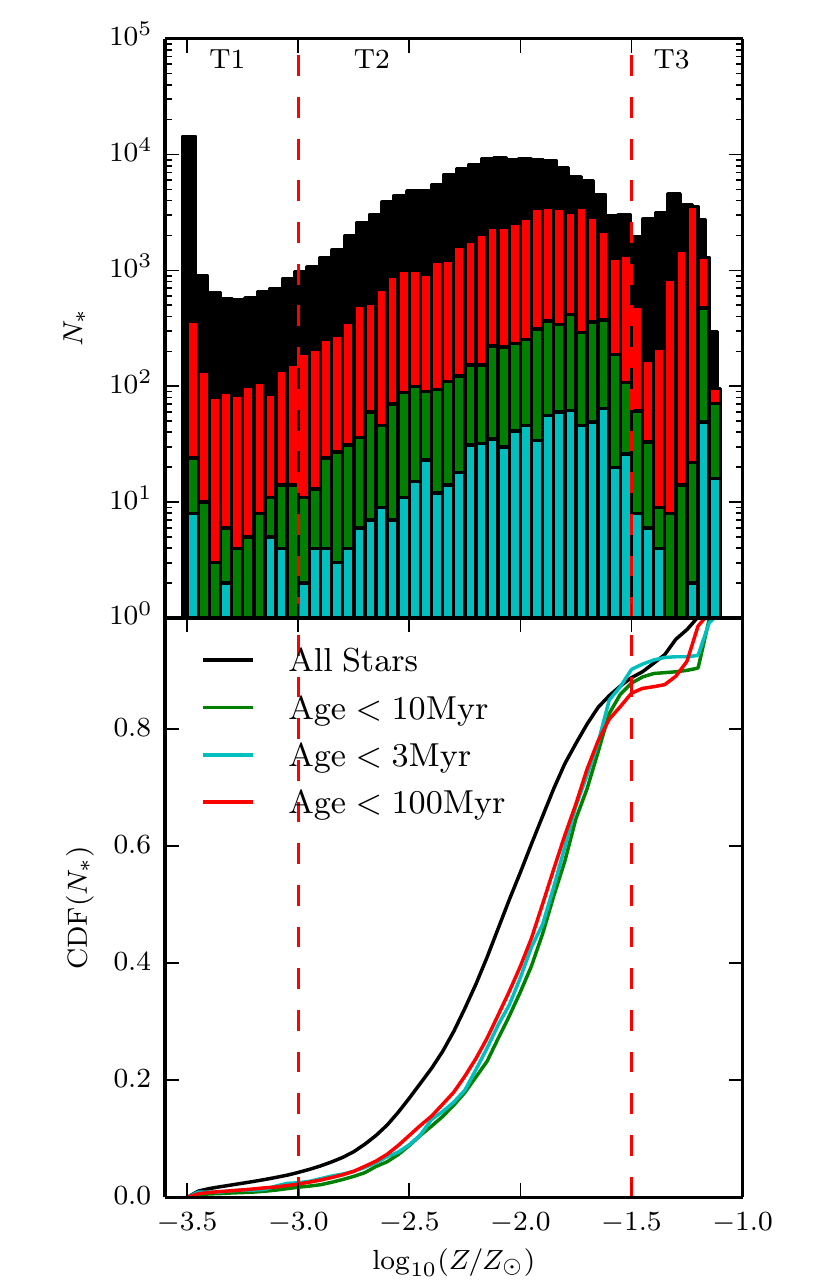}}
\caption{{\it Top}. Histograms of the number of star particles at a given metallicity at $z=6$.  Black shows all star particles while red, green, and cyan show stars less than 100~Myr, 10~Myr, and 3~Myr old, respectively.  The vertical dashed red lines represent the intervals for the three tracer classes T1, T2, and T3.  There is a strong peak at $Z=10^{-3.5}Z_{\odot}$ which is the metallicity floor of the simulation. {\it Bottom}.  Cumulative distribution functions of stellar metallicities in the simulation at $z=6$ split between all stars in black and stars less than 100~Myr, 10~Myr, and 3~Myr old in red, green, and cyan, respectively.   Approximately $5\%$ of the star particles are in T1, $\sim85\%$ of the star particles are in T2, and $\sim10\%$ of the star particles are in T3 at $z=6$.  The distribution is similar for the young stars although the highest metallicity bin contains $\sim14\%$ of all stars with age$<$100~Myr.  The CDF is skewed to higher metallicity for the younger generations of stars compared to all stars as expected.}
\label{metalhist}
\end{figure}

In the top panel of Figure~\ref{metalhist}, we plot a histogram of star particle metallicities in our simulation at $z=6$.  There is a noticeable peak at $Z=10^{-3.5}Z_{\odot}$, which is the value of the initial gas metallicity (we have assumed a Solar metal mass fraction of $Z_{\odot}=0.02$).  These represent the first generation of stars to form in a given halo in the simulation.  In order to capture the contribution from these stars, we set the first tracer bin to represent all photons that are emitted by stars with metallicity $Z<10^{-3}Z_{\odot}$.  The remaining stars show a bimodal distribution\footnote{This is partially due to having one very massive system in the box.} and therefore we place an empirically motivated cutoff at $Z\leq10^{-1.5}Z_{\odot}$, which splits the second and third tracer bins.  

The highest metallicity stars are generally present in the highest mass galaxies in the simulation, and we find that these systems exhibit metallicity gradients\footnote{The distribution of metals in the system is very dependent on stellar feedback \citep[e.g.,][]{Keating2016} and thus the convolution between metallicity and stellar galactocentric radius should be interpreted with caution as we have only tested one SN feedback algorithm.}.  Therefore, metallicity is a complicated convolution of halo mass with the location of the star particle in the halo.  Similarly, the lower metallicity stars emit more ionising photons than the higher metallicity stars which adds an additional complication.  Stars with lower metallicity also tend to have higher temperatures which leads to a harder spectrum.  Thus the metallicity of the stars that reionize the Universe will also shape the temperature evolution of the IGM.  Finally, while not explicitly used in our feedback scheme, it is expected that the momentum injection from SN is weakly dependent on metallicity \citep{Thornton1998}.  This could change the escape fraction for a momentum based feedback scheme such as the one presented in \cite{Kimm2014}.  Our  photon tracer algorithm can capture all of these effects. 

In the top panel of the middle column of Figure~\ref{combined_stats}, we plot the fractional contribution of each metal tracer bin to the instantaneous luminosity as a function of redshift.  At early times ($z\gtrsim13$) the lowest metallicity stars dominate the instantaneous luminosity.  However, at later times ($z\lesssim12$), systems become enriched with metals and the intermediate metallicity stars contribute most to the instantaneous luminosity.  

In the middle panel of the middle column of Figure~\ref{combined_stats}, we show the fractional contribution of each stellar metallicity tracer bin to the volume-weighted hydrogen photoionisation rate as a function of redshift in the ionised regions.  Following the luminosity, at the highest redshifts, $\Gamma_{\rm HI}$ is dominated by stars that have metallicities at the initial metallicity floor of the simulation.  The relative contributions fluctuate as a few metal enriched stars form at high redshift.  Once most systems become minimally enriched, the contribution from the lowest metallicity systems begins to decline.  This occurs in our simulation at $z\sim12$.

By $z\sim8$, there is a noticeable contribution from the highest metallicity stars to the global photoionisation rate in the simulation because systems have become significantly enriched.  At $z=6$, the photoionisation rate is distributed as $\sim6.3\%$ for the lowest metallicity stars, $\sim68.5\%$ for the middle metallicity stars, and $\sim25.2\%$ for the highest metallicity stars.  By instantaneous luminosity (averaged over the past 10Myr), the lowest metallicity, intermediate metallicity, and highest metallicity stars are contributing $\sim4\%$, $\sim69\%$, and $\sim27\%$ to this value, respectively.  
Similar to what we found for the mass tracers, we find that the lowest metallicity stars contribute more to the UV background at $z=6$ than expected from their luminosity.  This implies that the lowest metallicity systems have the highest escape fractions.  In the bottom panel of the middle column in Figure~\ref{combined_stats}, we plot the average escape fractions for each stellar metallicity tracer bin.  Over the course of reionization, the lowest, intermediate, and highest metallicity stars have mean escape fractions of 17.1\%, 11.5\%, and 15.9\%, respectively.

It is well established that there exists a mass-metallicity relation so that the lowest mass galaxies are also expected to have the lowest metallicity stars \citep[e.g.,][]{Tremonti2004}.  By comparing the top and middle rows in Figure~\ref{z8mm}, one can see that the red bubbles located in the top row, which represent the ionised regions where the UV background is dominated by the lowest mass galaxies, also appear in the middle row that shows the same quantity for the lowest metallicity systems.  Similarly, the regions where the UV background is dominated by the highest metallicity stars are also co-spatial with those regions where the UV background is dominated by the highest mass haloes.  

\begin{figure}
\centerline{\includegraphics[scale=1.0]{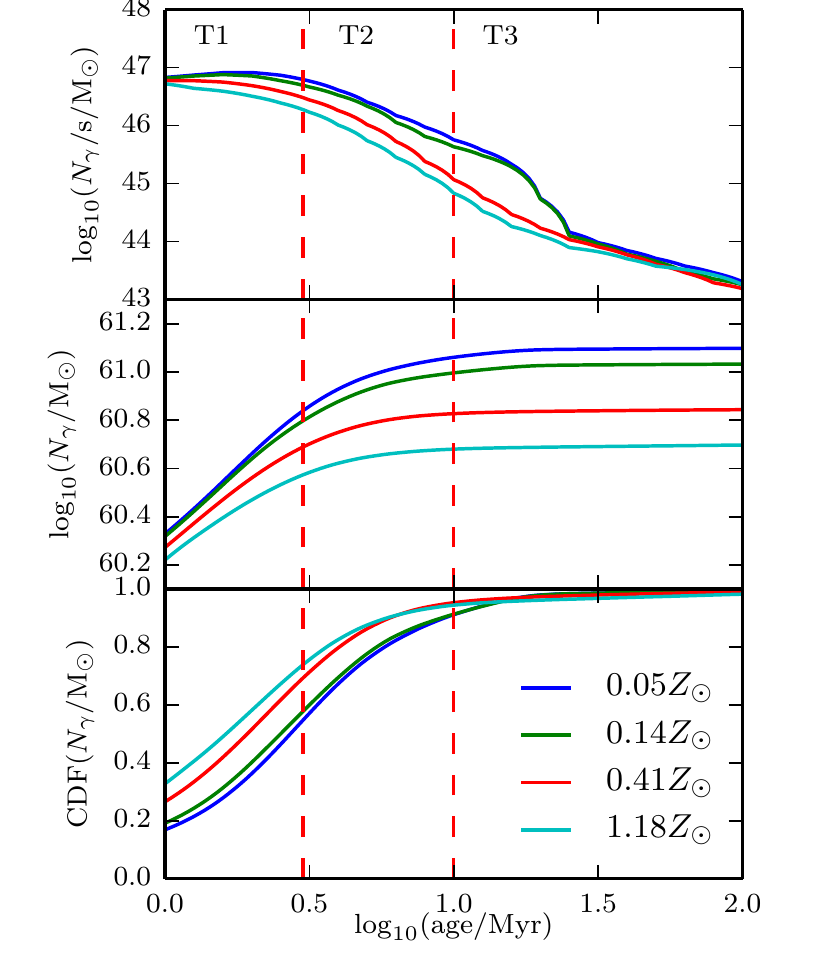}}
\caption{{\it Top}. Stellar luminosities per solar mass as a function of age from the {\small BPASS300} model in the hydrogen ionising bin are shown for different metallicity stellar populations as the curves of different colours, as indicated in the legend.  The vertical dashed red lines represent the intervals for the three tracer classes T1, T2, and T3.  The peak luminosity occurs in T1 for all metallicities.  {\it Middle}.  Cumulative number of photons emitted per solar mass as a function of age.  {\it Bottom}. Cumulative distribution function of the total number of photons emitted per solar mass as a function of age.  For the lowest metallicity bin, $\sim55\%$ of photons are emitted in T1, $\sim36\%$ of photons are emitted in T2, and $\sim9\%$ of photons are emitted in T3.  For higher metallicity systems, ionising photons are emitted earlier in the lifetime of the star. }
\label{SEDplot}
\end{figure}

\subsection{Tracing Age}
Unlike stellar metallicity, which is extremely dynamic as a function of time and depends on previous generations of star particles as well as the mass growth history of individual galaxies, stellar age is much less dependent on galaxy mass and affects all star particles equally.  In Figure~\ref{SEDplot}, we show the instantaneous luminosities, cumulative number of photons emitted, and cumulative distribution function of photons emitted as a function of stellar age for different metallicity stars in the {\small BPASS300} model.  For the lowest metallicity bin, which represents the bulk of the stars in our simulation at $z=6$, the peak luminosity occurs at an age less than 3Myr.  We have chosen this age as the cutoff of our first tracer bin as it represents roughly half of all of the ionising photons emitted by the star (at low metallicity).

For our chosen stellar feedback model, there is a delay of 10Myr between star particle formation and SN.  We expect that the SN may help clear low density, ionised channels where the ionising photons can escape out of the galaxy and into the IGM \citep{Kimm2014,Wise2014,Trebitsch2017,Rosdahl2018}.  Thus we place another tracer bin cutoff at an age of 10Myr to separate the photons emitted before and after a SN.  In total, only $\sim9\%$ of all ionising photons are emitted after an age of 10Myr for the lowest metallicity star particles.  This decreases mildly with increasing metallicity (see the bottom panel of Figure~\ref{SEDplot}).

In the top right panel of Figure~\ref{combined_stats}, we show the fractional contribution of each stellar age tracer bin to the instantaneous luminosity in the simulation as a function of redshift.  The ratios of these quantities between the different stellar age tracer groups are exactly what is expected when examining the CDF of the total number of photons emitted in the bottom panel of Figure~\ref{SEDplot}.  

In the centre panel of the third column of Figure~\ref{combined_stats}, we show the fractional contribution of each stellar age tracer bin to the volume-weighted hydrogen photoionisation rate in ionised regions as a function of redshift.  Across the entire history of reionization, the photons emitted by stars with ages between $3-10$Myr are dominating the photoionisation rate.  Interestingly, this contribution decreases from $\sim80\%$ to $\sim40\%$ with decreasing redshift from $z=18-9$ while the contribution from the photons emitted by the youngest stars steadily increases during this period.  The photons from the youngest stars likely encounter the highest quantity of neutral gas and hence they are absorbed locally.  This is expected to create channels (via photoionisation and photoheating) for photons emitted by slightly older stars to escape into the IGM. Thus, it is the middle aged stars that dominate the photoionisation rate.  Eventually SN will explode and create additional channels for ionising radiation to escape. The increasing contribution from the photons emitted by the youngest stars indicates that the escape fraction for this population of photons is rising with decreasing redshift.  This is evident in the bottom right panel of Figure~\ref{combined_stats}.  The contribution from the old age stars to the total photoionisation rate is clearly much greater than the $\sim9\%$ expected by simply integrating the SED over time, demonstrating that relatively few of these photons are absorbed locally around the stars.  The escape fraction of this population is extremely high throughout the course of the simulation and remains within the range of $40\%-60\%$.

At $z\sim9$, there is a sharp drop in the contribution to the photoionisation rate from stars younger than 3Myr while at the same time, the contribution from stars with age between $3-10$Myr increases by $\sim20\%$.  We have marked this event with a dashed vertical line in the middle panel of the right column of Figure~\ref{combined_stats}.  This corresponds to the time where we introduce an additional refinement level into the AMR hierarchy in the simulation (to maintain a roughly constant physical resolution).  Introducing an additional level in the grid affects the star formation.  The luminosity of young stars decreases sharply after this event before steeply rising again (see the top right panel of Figure~\ref{combined_stats2}).  Because the photons that remain present in the IGM are from intermediate and old aged stars, the relative contribution of the young stars to $\Gamma_{\rm HI}$ decreases as few are emitted immediately following the addition of a refinement level. 

We must keep in mind that the escape fraction from different aged stars is likely very sensitive to the feedback model chosen for the simulation.  We have used a fixed value of 10Myr after the formation of a star before energy is injected to model SN.  This is rather unrealistic as it is well established that the actual SN rates vary as a function of time \citep[e.g.,][]{Leitherer1999}.  The earliest SN explode only a few Myr after formation and by neglecting this in our simulation, we are artificially suppressing the ionised channels that may form before 10Myr and change the escape fractions of photons emitted by the youngest stars.  We once again expect that the qualitative behaviour holds that the photons emitted by the middle aged stars will have a higher relative escape fraction compared to the photons emitted by the youngest stars; however, the exact ratio of these two escape fractions is likely to be very dependent on the way that SN feedback is implemented in the simulation.  For this reason, we aim to study how SN feedback affects the escape fraction of different populations of photons in future work.

\subsection{Comparison of $\Gamma_{\rm HI}$ with $V_f$}
\label{gammavf}
\begin{figure}
\centerline{\includegraphics[scale=1.0,trim={0 1cm 0 0cm},clip]{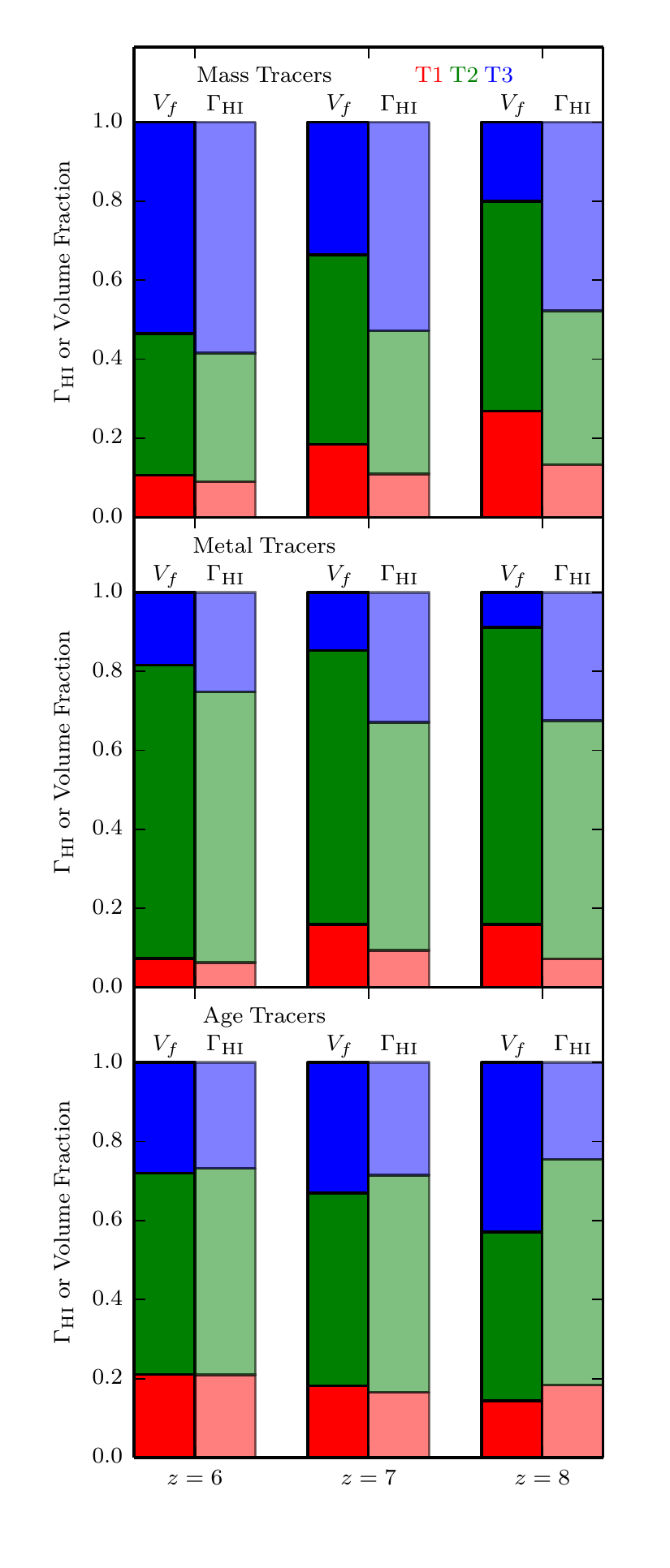}}
\caption{Comparison of the fractional contribution of each of the different photon tracers to the volume-weighted photoionisation rate in the ionised regions (light) with the fraction of the ionised volume filled by photons from each of the different sources (dark).  In all three panels, T1 (low mass haloes, low metallicity stars, young stars), T2 (intermediate mass haloes, intermediate metallicity stars, middle aged stars), and T3 (high mass haloes, high metallicity stars, old stars) are shown in red, green and blue, respectively.  The top, middle, and bottom panels show the results for the mass, metal, and age tracers, respectively.  At $z=6$, when the box is completely ionised and the photoionisation background is in place, the two quantities agree.  At higher redshifts, the quantities diverge as the photoionisation rate changes in bubbles where the UV background is set by different types of sources.}
\label{vgrat}
\end{figure}

Throughout our analysis, we focused on measuring the contribution of each of the different sources of ionising photons to the globally averaged photoionisation rate in ionised regions.  Instead of describing which sources are providing the bulk of the photons contributing to the photoionisation rate, we can measure the volume of the Universe filled with photons emitted by different sources.  

In Figure~\ref{vgrat}, we quantify the volume filling factors of each of the different sources for each of the different tracers at three redshifts ($z=6$, $z=7$, and $z=8$).  Beginning with the mass tracers, at $z=8$, the volume filling fraction of ionised regions dominated by the most massive galaxies is far less than the contribution these highest mass galaxies have to the total photoionisation rate.  This implies that photons emitted by intermediate mass haloes fill most of the volume of ionised regions but the photoionisation rate in HII bubbles filled by photons from the highest mass systems are significantly higher than average.  This is consistent with our analysis of Figure~\ref{gamrho_mass} that the photoionisation rate is higher in the regions surrounding massive galaxies.  We expect that the volume filling fraction from high mass galaxies will increase as the HII bubbles surround these systems grow.  In the top panel of Figure~\ref{vgrat}, we can see that the length of the dark blue bars increases from $z=8$ to $z=6$ indicating that the photons emitted by highest mass galaxies are progressively filling more of the volume.  At $z=6$, when the simulation volume is completely ionised, the volume filling fraction agrees well with the fraction contributed to the global photoionisation rate.  This is because all of the HII bubbled have merged and any region of the simulation can see photons from all sources.  This behaviour is expected to hold regardless of the quantity we are tracing.  

As we have shown previously, stellar metallicity is a reasonably good proxy for halo mass and in the middle panel of Figure~\ref{vgrat}, we show that at $z=8$, the photoionisation rate has a strong contribution from the highest metallicity stars while the ionised volume is mostly filled by photons emitted by middle and lower metallicity stars.  Hence there are fluctuations in $\Gamma_{\rm HI}$ driven by different metallicity stars.

\section{Caveats}
\label{cavs}
There are still many caveats that should be kept in mind when assessing the results from this first demonstration of our algorithm.  These range from the moderate mass and spatial resolution as well as the moderate box size of the simulations to the uncertainties associated with the use of a particular set of subgrid models for star formation and stellar feedback.  We should also emphasise once more that the simulations presented here do not yet properly resolve the multiphase ISM, and in particular, the ionised channels through which LyC photons escape into the IGM.  Similarly the stellar feedback model used in the simulations is rather simplistic.  Because of the assumed delayed cooling, much of the dense gas remains hot, which results in an unrealistic temperature-density distribution.  In reality, SNe will drive the gas to lower densities and higher temperatures, which will change the escape of photons into the IGM.  Furthermore, our star particles represent entire stellar clusters and we do not properly model the scatter in individual birth times for stars in a real cluster.  Thus our claims on which age stars contribute to the UV background is subject to this assumption.  If this scatter is larger than the time it takes until the first stars in the cluster explode via SN, the contribution from young stars to the UV background would increase.

One of the most important open questions regarding reionization is whether galaxies are the dominant source of LyC photons and which mass range of galaxies reionized the Universe.  In order to answer this question more definitively, a significantly larger volume will need to be simulated at higher resolution. Our algorithm is very well suited for this type of work, but such simulations will come at a considerable computational cost.  Note also that with the M1 closure, a detailed spatial analysis of which photons contribute to the UV background in specific regions becomes problematic after the percolation of HII regions. A different radiative transfer technique allowing a better mixing of radiation from different sources may be needed to advance this type of work past percolation.

\section{Discussion \& Conclusions}
\label{dc}
We have presented an implementation of a new photon tracer algorithm designed to track the contribution of various source populations to the reionization history of the Universe in {\small RAMSES-RT}.  Our new algorithm  avoids the need to explicitly measure the escape fraction of individual sources in the simulation at all redshifts in order to measure their contribution to the metagalactic UV background.  We have demonstrated three applications for our algorithm in a cosmological simulation that agrees well with observations in terms of stellar mass-halo mass ratio, reionization history, and photoionisation rate. We explicitly tracked the contribution to the photoionisation rate from sources classified according to halo mass, stellar metallicity, and stellar age.  

We have demonstrated here the following capabilities of our new algorithm.
\begin{enumerate}
\item We measure the contribution of various source populations to the UV background at any given redshift.  This allowed us to study how these contributions vary spatially as well as temporally and to determine which sources dominate the UV background in different environments at different times.  It further allowed us to study the origin of the fluctuations of  $\Gamma_{\rm HI}$ during and after reionization.
\item We used our new algorithm to measure the globally averaged escape fractions of different  classes of sources  on-the-fly in the simulation as a function of redshift.  This will be particularly useful when running simulations with higher resolution than presented here which are required to resolve the ISM better.
\item We have presented an algorithm that allows us to measure the contribution of various source populations to the ionisation fraction of the gas. This will allow us to study which sources reionized the Universe as a function of redshift (to be presented in a follow-up paper).  This can be used to study spatial as well as temporal fluctuations of the contribution of different source population to the ionisation fraction. 
\end{enumerate}

We have identified  a number of interesting trends in our simulations that are likely to hold for higher resolution simulations:
\begin{enumerate}
\item The highest mass systems ($M>10^{10}$M$_{\odot}h^{-1}$) in our simulations dominate the UV background at $z=6$ while the escape fraction decreases with increasing halo mass.
\item HII bubbles ionised by the highest mass galaxies tend to have higher photoionisation rates than bubbles ionised by lower mass galaxies ($M<10^{10}$M$_{\odot}h^{-1}$) before overlap occurs.  This may lead to spatial variations of the properties of dwarf galaxies that are forming in different bubbles.
\item At early times the photoionisation rate in our simulations is dominated by the lowest metallicity stars, but as the overall metallicity increases, their contribution decreases.  Intermediate metallicity stars ($10^{-3}Z_{\odot}<Z\leq10^{-1.5}Z_{\odot}$) dominate the photoionisation rate at $z=6$.
\item Photons emitted by stars in the age range 3-10 Myr dominate the photoionisation rate across all redshifts.  The photons from the youngest stars create low opacity channels that allow photons emitted by slightly older stars to penetrate efficiently into the IGM.  As a result, the oldest stars emitting ionising photons have very high escape fractions of $40-60\%$.
\end{enumerate}

We have concentrated here on three quantities for the origin of ionising photons: halo mass, stellar metallicity, and stellar age.  The algorithm is, however, adaptable to track any quantity that can be measured in the simulation and we anticipate future applications to include directly measuring the contribution from, for example, AGN, Pop II, and Pop III stars, the stellar mass fraction of a galaxy, the stellar galactocentric radius at formation, the efficiency at which star particles are forming, and the gas density at the star formation location.  With photon tracers, one can also study the properties of haloes residing in bubbles ionised by different sources.  This will give insight into whether quantities such as the galaxy filtering mass  depend on the source that reionized the region or whether the properties of Ly$\alpha$ emitters, damped Ly$\alpha$ systems, or Lyman limit systems differ as well.  Similarly, our algorithm may help to better understand temperature fluctuations in the IGM.  There will also be many applications that go beyond studying reionization.  Our algorithm is, for example, well suited to track the momentum transfer of photons from different sources inside of a galaxy and should allow for determination of which sources are providing the majority of the radiation pressure.  Photon tracers should thus  become a very useful tool in the studies of the reionization epoch with radiative transfer simulations and beyond.

\section*{Acknowledgements}
HK thanks Girish Kulkarni for useful discussions and comments on this manuscript.  This work made considerable use of the open source analysis software {\small PYNBODY} \citep{pynbody}. HK's thanks Foundation Boustany, the Cambridge Overseas Trust, and the Isaac Newton Studentship.  HK also thanks Brasenose College and the support of the Nicholas Kurti Junior Fellowship as well as the Beecroft Fellowship. Support by ERC Advanced Grant 320596 ``The Emergence of Structure during the Epoch of reionization'' is gratefully acknowledged.  DS acknowledges support by STFC and ERC Starting Grant 638707 ``Black holes and their host galaxies: coevolution across cosmic time''.  TK acknowledges support by the National Research Foundation of Korea to the Center for Galaxy Evolution Research (No. 2017R1A5A1070354) and in part by the Yonsei University Future-leading Research Initiative of 2017 (RMS2-2017-22-0150).  JR and JB acknowledge support from the ORAGE project from the Agence Nationale de la Recherche under grand ANR-14-CE33-0016-03.

This work was performed using the DiRAC/Darwin Supercomputer hosted by the University of Cambridge High Performance Computing Service (http://www.hpc.cam.ac.uk/), provided by Dell Inc. using Strategic Research Infrastructure Funding from the Higher Education Funding Council for England and funding from the Science and Technology Facilities Council. 

This work used the DiRAC Complexity system, operated by the University of Leicester IT Services, which forms part of the STFC DiRAC HPC Facility (www.dirac.ac.uk). This equipment is funded by BIS National E-Infrastructure capital grant ST/K000373/1 and  STFC DiRAC Operations grant ST/K0003259/1. DiRAC is part of the National E-Infrastructure

Furthermore, this work used the DiRAC Data Centric system at Durham University, operated by the Institute for Computational Cosmology on behalf of the STFR DiRAC HPC Facility (www.dirac.ac.uk).  This equipment was funded by the BIS National E-infrastructure capital grant ST/K00042X/1, STFC capital grant ST/K00087X/1, DiRAC operations grant ST/K003267/1 and Durham University.  Dirac is part of the National E-Infrastructure.

\appendix
\section{Absolute Luminosities, Cumulative Luminosities, and Absolute $\Gamma_{\rm HI}$ for Each Tracer Group}
\label{luminfo}
\begin{figure*}
\centerline{\includegraphics[scale=1.0,trim={0 1.5cm 0 1cm},clip]{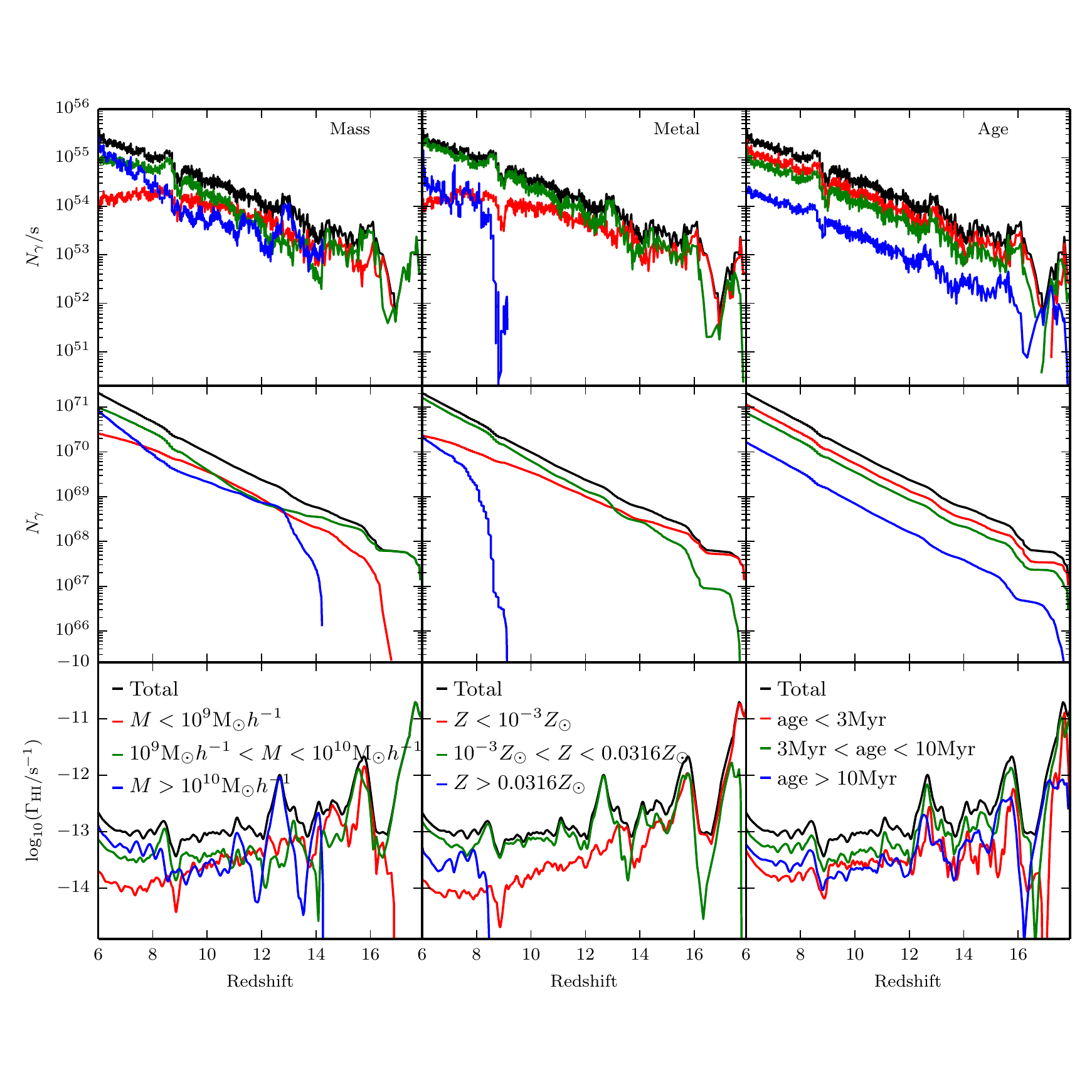}}
\caption{{\it Top}. Instantaneous luminosity of each tracer group as a function of redshift.  The left, centre, and right panels show the results for the halo mass tracers, stellar metallicity tracers, and stellar age tracers, respectively.  The red, green, and blue lines represent sources in T1 (low mass haloes, low metallicity stars, young stars), T2 (intermediate mass haloes, intermediate metallicity stars, middle aged stars), and T3 (high mass haloes, high metallicity stars, old stars), respectively. {\it Middle}.  Cumulative number of photons emitted for each tracer group as a function of redshift.  {\it Bottom}. $\Gamma_{\rm HI}$ for each tracer group as a function of redshift. }
\label{combined_stats2}
\end{figure*}

In Figure~\ref{combined_stats2}, we show the absolute instantaneous luminosities for each tracer group as a function of redshift in the top row, the cumulative number of photons emitted by each tracer group as a function of redshift in the middle row, and the absolute $\Gamma_{\rm HI}$ for each tracer group in the bottom row.

\section{Speed of Light Convergence}
\label{SLC}
\begin{figure}
\centerline{\includegraphics[scale=1.0]{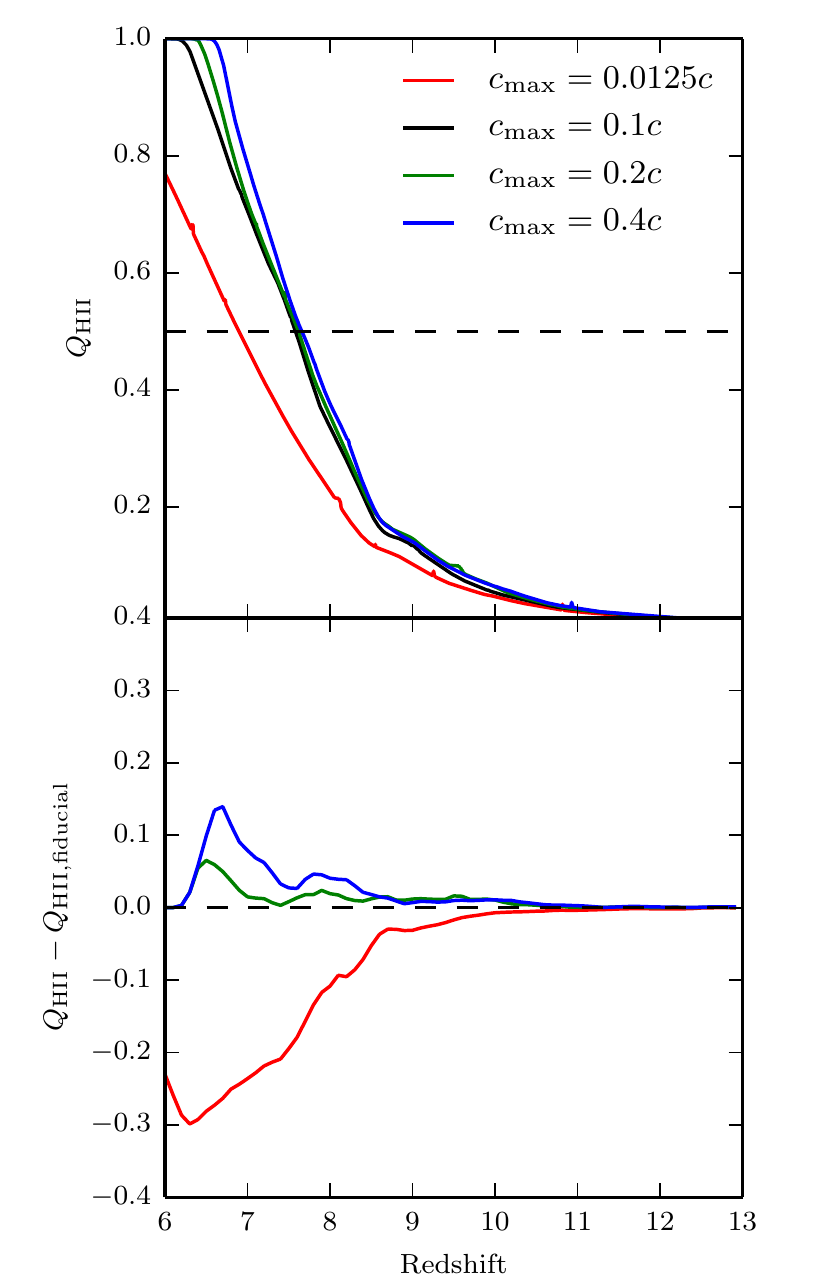}}
\caption{{\it Top}. The volume filling factor of ionised hydrogen as a function of redshift in our simulation for various different values for the speed of light.  The three simulations that use the VSLA algorithm are well converged until percolation sets in at a volume filling fraction of $Q_{\rm HII}\sim0.5$ indicated as the dashed horizontal line.  When this occurs, the simulations that use the faster value of the speed of light ionise the box slightly faster.  {\it Bottom.} Difference in $Q_{\rm HII}$ between the test simulation and the fiducial simulation used in the main text.  The maximum difference in the ionised hydrogen fraction between the three VSLA runs is only $\sim10\%$ which occurs during percolation.  The run which uses $c_{\rm sim}=0.0125c$ on all AMR levels differs from the fiducial run long before percolation and under predicts the ionisation fraction by up to $\sim30\%$.  For a larger box with more luminous sources, we expect this difference to grow.}
\label{cconv}
\end{figure}

Our fiducial simulation in this work has adopted a maximum speed of light of $0.1c$, which is the value of $c_{\rm sim}$ set on the base grid of the AMR hierarchy.  We have run three additional simulations: 1) we assume $c_{\rm sim}=0.0125c$ on all levels, 2) we increase the maximum speed of light in the simulation by a factor of 2, and 3) we increase the maximum speed of light by a factor of 4.  In the top panel of Figure~\ref{cconv}, we show the volume filling factor of ionised hydrogen for each of the three supplemental models and compare this with the fiducial model used in the main text.  As expected, the red line, which is the model that uses the slowest value of the speed of light, lags behind the other solutions at all relevant redshifts \citep{Bauer2015,Katz2017}.  In some cases, the simulation box is up to 30\% less ionised in this run compared to the fiducial simulation.  The other two simulations are reasonably well converged with our fiducial model (in terms of the volume filling factor of ionised hydrogen) up to the point when the simulation is $\sim50\%$ ionised.  At this point, percolation starts to set in and the photons can propagate long distances in the optically thin regime.  Hence the models using the faster value of the speed of light begin to deviate.  In terms of reionization history, the deviation is rather modest and we opt to use $0.1c$ as our maximum value for computational efficiency.  The maximum deviation between the fiducial model and the run with $c_{\rm max}=0.4c$ is only $\sim20\%$ while the mean deviation remains around $\sim10\%$ over the course of the simulation.

In terms of computation time, our fiducial model is only 35.6\% slower than the $c_{\rm max}=0.0125c$ run while the $c_{\rm max}=0.2c$ and $c_{\rm max}=0.4c$ runs are 80\% and 171\% slower, respectively.  The substantial reduction in run time is a strong motivation for the use of the variable speed of light approximation.  Note that the timings presented here are significantly improved over those presented in the Appendix of \cite{Katz2017}.

Clearly, the runs that use a faster speed of light on the base grid reionize the box slightly faster.  The question we would like to address here is how is the evolution of the photoionisation rate affected for a given reionization history that is consistent with observations? Nearly all of our analysis of the photon tracers in the main text was dependent on the photoionisation rate from different classes of sources.  Thus it is this parameter which we aim to understand in terms of convergence with the speed of light.  

\begin{figure}
\centerline{\includegraphics[scale=1.0,trim={0 0.8cm 0 1.3cm},clip]{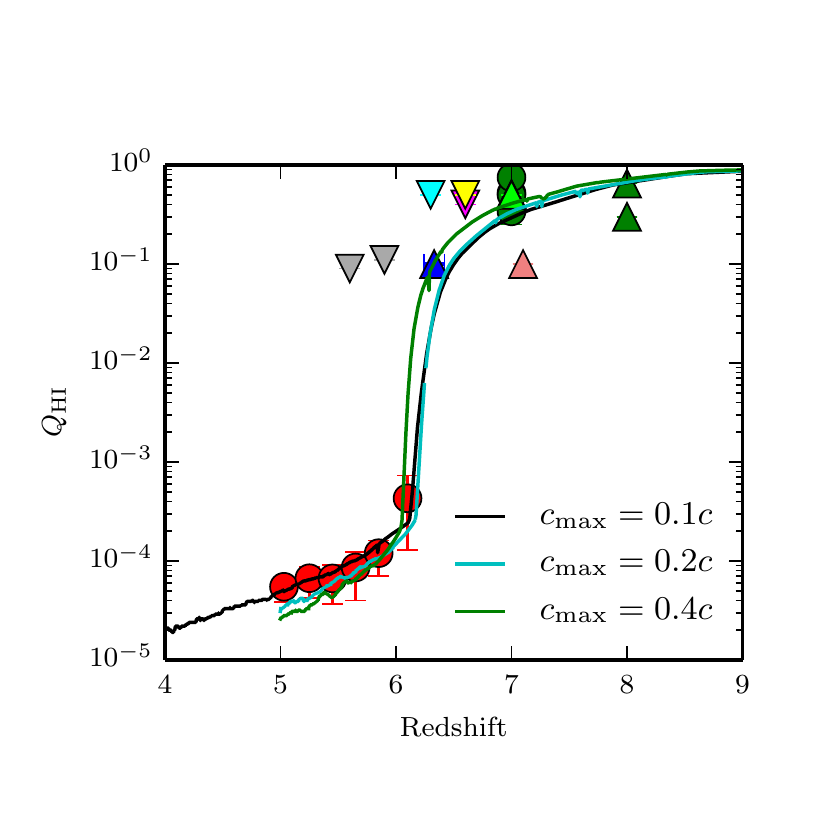}}
\centerline{\includegraphics[scale=1.0,trim={0 0.9cm 0 1.3cm},clip]{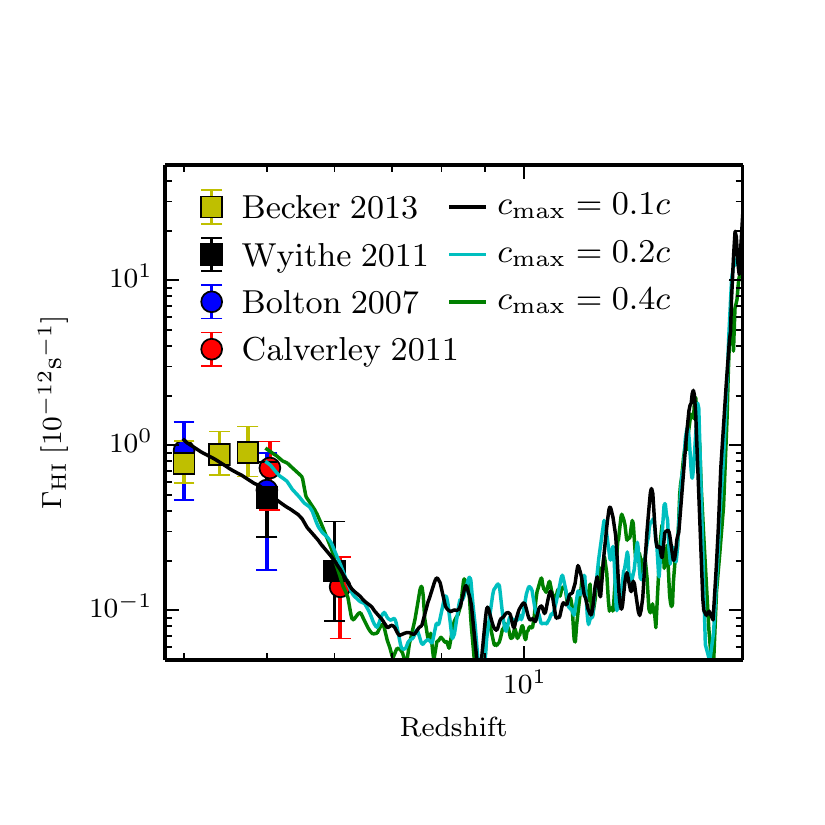}}
\caption{{\it Top}. The volume filling factor of neutral hydrogen as a function of redshift for our fiducial simulation is shown as the solid black line.  This is compared with two additional simulations that use an increased speed of light on the base grid by a factor of two and four which are shown in cyan and green, respectively.  We compare to various observations of this value compiled by \protect\cite{Bouwens2015}.  The reionization histories agree very well between the three models and likewise are consistent with observations.  {\it Bottom}.  The volume-weighted hydrogen photoionisation rate in ionised regions as a function of redshift for our fiducial model is shown as the solid black line.  The same values are computed for the twice enhanced and four times enhanced speed of light in cyan and green, respectively.  We define an ionised region as those cells which are at least 50\% ionised.  Observational constraints are shown as the data points and have been adopted from \protect\cite{Bolton2007,Wyithe2011,Calverley2011,Becker2013}.  All simulations agree remarkably well up to $z\sim6$ which justifies our use of $c_{\rm sim}=0.1c$ for the fiducial model.  Beyond $z=6$, the models with the faster speed of light have a volume-weighted photoionisation rate that increases much more rapidly.}
\label{gamconv}
\end{figure}

In order to test how speed of light impacts the photoionisation rate, we modify $f_{\rm lum}$ for the $c_{\rm max}=0.2c$ and $c_{\rm max}=0.4c$ runs so that they have the same reionization histories as the fiducial model.  This amounts to decreasing $f_{\rm lum}$ to slow down reionization as these models reionize too quickly as shown in the top panel of Figure~\ref{cconv}.  For the $c_{\rm max}=0.2c$ simulation, we set $f_{\rm lum}=1.102$ while for the $c_{\rm max}=0.4c$ simulation, we set $f_{\rm lum}=1.011$.  Recall that for the fiducial simulation $f_{\rm lum}=1.193$ and thus by decreasing $f_{\rm lum}$ by 0.091 every time the speed of light on the base grid is increased by a factor of two, we can achieve a reasonably converged reionization history. 

In the top panel of Figure~\ref{gamconv}, we plot the reionization histories of all three simulations.  Note that the runs with $c_{\rm max}=0.4c$ and $c_{\rm max}=0.2c$ were both stopped at $z=5$.  The agreement between the $c_{\rm max}=0.2c$ simulation and our fiducial model is remarkably good.   The lines overlap over the entirety of the epoch of reionization.  The $c_{\rm max}=0.4c$ simulation reionizes at a slightly later redshift; although, we should consider the log scaling of the y-axis in this plot and the actual percentage difference in terms of the volume filling factor of neutral hydrogen is extremely small.  With the reionization histories calibrated, we can now examine the photoionisation rates.  

\begin{figure}
\centerline{\includegraphics[scale=1.0,trim={0 0.8cm 0 1.1cm},clip]{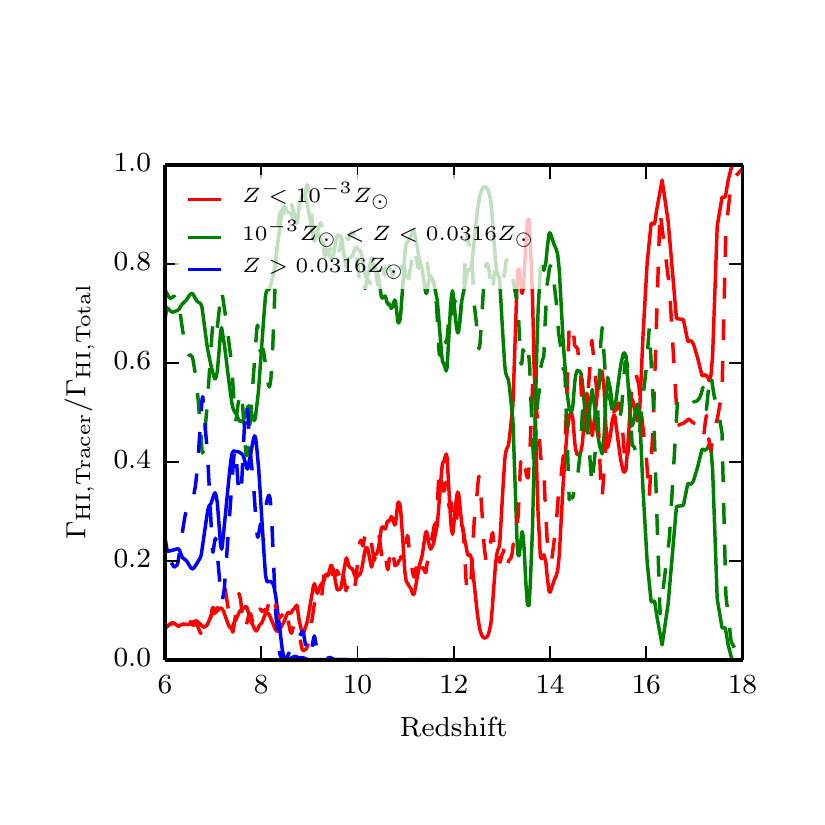}}
\caption{Comparison of the fractional contribution of the different metal tracer classes to the volume-weighted mean photoionisation rate in ionised regions as a function of redshift.  T1, T2, and T3 are shown in red, green, and blue, respectively.  Our fiducial model is shown as the solid lines while the comparison $c_{\rm max}=0.2c$ simulation is shown as dashed lines.  The agreement between the two simulations is reasonably good across the entire redshift interval suggesting that the fractional contributions of each photon tracer are rather insensitive to the choice of speed of light on the base grid.}
\label{gamconvmet}
\end{figure}

In the bottom panel of Figure~\ref{gamconv}, we plot the volume-weighted photoionisation rate of neutral hydrogen in ionised regions as a function of redshift for the three simulations.  Once again, we see very good agreement between all three models down to $z=6$, which is the final redshift we study using the photon tracers.  The convergence between reionization histories and photoionisation rates suggests that we are justified in using $c_{\rm sim}=0.1$ as our fiducial model, as long as we only study $z\geq6$.  Note that $f_{\rm lum}$ is constant for all sources so this should not change the relative contribution of each tracer class.  However, since there are fewer overall photons being emitted in the runs with the faster values of the speed of light, there may be a slight change in the contribution of each tracer class to the global photoionisation rate due to a different escape fraction out of the galaxies.  To test whether this was significant, we have run the  $c_{\rm max}=0.2c$ simulation with metal tracers and compare to our fiducial model.  In Figure~\ref{gamconvmet}, we plot the fractional contribution of each of the different metal tracer groups for the fiducial model (solid lines) compared to the simulation with $c_{\rm max}=0.2c$ (dashed lines).  Qualitatively, the two sets of simulations agree extremely well across the entire relevant redshift range as the two sets of lines track each other very well.  Both sets of simulations converge nicely to the same values and thus we can be reasonably confident that the tracer fractions are converged, independent of the speed of light on the base grid, as long as the reionization histories are kept constant.

The largest difference we see between the simulations occurs in the post-reionization epoch.  At $z<6$, we see that the photoionisation rates in the runs with a faster value of the speed of light on the base grid increase much more rapidly compared to the fiducial model.  This is because the photons are now propagating in the optically thin regime and hence  reducing the speed of light causes the photon fronts to lag.  From the top panel of Figure~\ref{gamconv}, we can see that the run with $c_{\rm max}=0.4c$ on the base grid ionises the simulation box slightly faster than the fiducial model in the post-reionization epoch.  In this regime, we must be more careful in interpreting the results with a reduced speed of light; however, at $z\geq6$ our results should be  robust.

\bibliographystyle{mnras}
\bibliography{Paper_I} 

\bsp	
\label{lastpage}
\end{document}